\documentstyle[aps,pre,array,multicol,epsfig,eqsecnum]{revtex}     

\begin{document}
\draft

\title{The coherent scattering function of the
       reptation model:\\ simulations compared to theory.}  
\author{Artur Baumg\"artner$^1$, Ute Ebert$^2$, Lothar Sch\"afer$^3$}
\address{$^1$ Institut f\"ur Festk\"orperforschung,
        Forschungszentrum J\"ulich, 52425 J\"ulich, Germany}
\address{$^2$ Centrum voor Wiskunde en Informatica, 
        P.O.\ Box 94079, 1090 GB Amsterdam, The Netherlands} 
\address{$^3$ Universit\"at Essen, Universit\"atsstr.\ 5, 45117 Essen, 
        Germany}
\date{slighty revised version from June 06, 2003 --- 
first submitted on March 11, 2002}
\maketitle

\begin{abstract}

We present results of Monte Carlo simulations measuring the coherent 
structure function of a chain moving through an ordered lattice of 
fixed topological obstacles. Our computer experiments use chains up 
to 320 beads and cover a large range of wave vectors and a time range 
exceeding the reptation time. For additional information we also measured 
the coherent structure function of internal pieces of the chain.

We compare our results $(i)$ to the predictions of the primitive chain model, 
$(ii)$ to an approximate form resulting from Rouse motion in a coiled tube, 
and $(iii)$ to our recent evaluation of the full reptation model. 
$(i)$ The primitive chain model can fit the data for times $t \agt 20 T_2$, 
where $T_2$ is the Rouse time of the chain. Besides some phenomenological 
amplitude factor this fit involves the reptation time $T_3$ as a second fit 
parameter. For the chain lengths measured, the asymptotic behavior 
$T_3 \sim N^3$ 
is not attained. $(ii)$ The model of Rouse motion in a tube, which we 
have criticized before on theoretical grounds, is shown to fail also 
on the purely phenomenological level. $(iii)$ Our evaluation of the full 
reptation model yields an excellent fit to the data for both total 
chains and internal pieces and for all wave vectors and all times, 
provided specific micro-structure effects of the MC-dynamics are 
negligible. Such micro-structure effects show up for wave vectors 
of the order of the inverse segment size and enforce the introduction 
of some phenomenological, wave vector dependent prefactor. For the 
dynamics of the total chain our data analysis based on the full 
reptation model shows the importance of tube length fluctuations. 
Universal (Rouse-type) internal relaxation, however, is unimportant. 
It can be observed only in the form of the diffusive 
motion of a short central subchain in the tube.

Finally we present a fit formula which in a large range of wave vectors 
and chain lengths reproduces the numerical results of our theory 
for the scattering from the total chain.   
\end{abstract} 

%\newpage

\begin{multicols}{2}

\section{Introduction}

Dynamical properties of dense polymer systems like melts or dense 
solutions often are analyzed within the framework of the reptation 
model \cite{Z1,Z2}. Reptation is a specific mechanism for the motion 
of a single tagged chain through an environment of other chains. 
It is based on the idea that the background chains act as impenetrable 
obstacles which confine the motion of the tagged chain to a tube roughly 
defined by its present configuration. The local motion of the inner parts 
of the chain is restricted to the diffusion of little wiggles of 
'spared length' along the tube. Globally the motion is driven by 
the chain ends, where wiggles are created or destroyed. Creation of 
a wiggle shortens the tube by its spared length, destruction prolongs 
the tube in some randomly choosen direction. In the long run this motion 
of the chain ends leads to the complete destruction of the original tube 
and to large scale diffusion of the chain.

Formulated in more precise terms, the reptation model deals with the 
stochastic motion of a flexible chain embedded in a fixed environment 
of obstacles which form the edges of a regular lattice in three dimensional 
space. In this work we present results for the coherent structure function 
measured in an extensive simulation of this model. The measured coherent 
structure functions of the total chain and of internal subchains are 
compared to the results of our recent analytical evaluation \cite{Z3} 
of the model. For the total chain there exist previous approximate 
theories based on reptation \cite{Z4,Z5}, which are included in the 
comparison. These theories do not treat the full dynamics of the model, 
but neglect so called `tube length fluctuations'. For typical chain 
lengths used in (computer- or physical-) experiments these fluctuations 
are known to yield important contributions, as has first been pointed 
out in Ref.\ \cite{Z6} in the context of an analysis of the viscosity.

As mentioned above, results of the reptation model generally are used 
to analyze data for systems like polymer melts \cite{Z7}, where the 
surrounding of a given chain certainly is far from forming an ordered 
time independent lattice of obstacles. Clearly the surrounding chains 
slowly move away, which leads to `constraint release' \cite{Z8}, 
an effect that becomes important \cite{Z7} outside the limit of 
asymptotically long chains. Also disorder in the distribution of obstacles 
might lead to fluctuations in the local tube diameter, thus affecting 
the local mobility of spared length. In this work we omit all such 
effects of the environment and study the coherent structure function 
of the pure reptation model, as described above. This is a necessary 
prerequisite for an analysis aiming at the separation of the different 
effects present in a real melt.

To illustrate the problem we now briefly recall some typical results 
\cite{Z1,Z2} of reptation, as established for very long chains. 
We concentrate on the motion of an internal segment, which from 
a theoretical point of view is the simplest quantity to discuss.

Simple as it is, the reptation scenario involves several time scales 
and leads to a rich phenomenology. It needs a microscopic time $T_{0}$ 
before the chain feels the existence of constraints due to its surrounding. 
Generally $T_{0}$ is taken as the Rouse time of a short subchain of $N_{e}$ 
segments: $T_{0} \sim N_{e}^{2}$, where the 'entanglement length' $N_{e}$ 
is choosen such that the coil diameter of the subchain is of the order of 
the diameter of the tube, which substitutes the surrounding chains. 
The second time scale $T_{2}$ is the relaxation time of the total chain 
in a fixed tube, i.e. the time a wiggle needs to diffuse over the whole 
chain. It depends on chain length $N$ as 
$T_{2} \sim T_{0} (N/N_{e})^{2} \sim N^{2}$ and thus behaves 
as the relaxation time of a free chain in the Rouse model. 
The longest scale $T_{3}$ is the 'reptation time'. It measures the 
time which the motion of the chain ends needs to completely destroy 
the initial tube. In the limit of long chains the reptation model 
predicts \cite{Z1} $T_{3} \sim (N/N_{e})^{3} T_{0}$.

For observables like the motion of individual segments the model 
yields asymptotic power laws, where the exponent depends on the time 
range. We here quote the results for the motion of the central segment 
$j = N/2$:  
\begin{equation}
\left\langle
\overline{\left(
{\bf r}_{N/2}(t) - {\bf r}_{N/2}(0)
\right)^{2}}
\right\rangle
\sim 
\left\{ 
\begin{array}{l@{\quad;\quad}l}
t^{1/4} & T_{0} \ll t \ll T_{2}\\
(t/N)^{1/2} & T_{2} \ll t \ll T_{3}\\
t/N^{2}      & T_{3} \ll t
\end{array}
\right.
\end{equation}

We use the bar to denote the dynamic average, i.e. the average over 
the motion of spared length. The pointed brackets stand for the average 
over all initial configurations. 

Considerable effort has been invested to check these predictions in 
simulations of melts, but the outcome to date is not conclusive 
\cite{Z7,Z9,Z10}. The $t^{1/2}$-regime has never been properly 
identified. (Note that crossover from a region 
$\langle \overline{({\bf r}_j(t) - {\bf r}_j(0))^2}\rangle 
\sim t^\alpha, \: \alpha < \frac{1}{2}$, to free diffusion easily 
can pretend the existence of a $t^{1/2}$-regime. What has to be 
demonstrated is the stability of this regime for a larger range 
of time and chain lengths.) Slowing down of segment motion in the range 
$T_0 < t < T_2$ is observed \cite{Z9}, with an effective exponent 
somewhere between 1/4 and 1/2. Only for some related observable, 
measuring the motion of a central segment relative to the center of mass, 
a $t^{1/4}$-behavior seems to be established \cite{Z10}. 
Real experiments do not measure $\langle \overline{({\bf r}_j(t) - 
{\bf r}_j(0))^2}\rangle$. However, a related quantity, 
the return-to-origin probability of a segment averaged over 
all segments, is measured in NMR-experiments. Here the equi\-valent 
to $t^{1/4}$-behavior has been found in Ref.\ \cite{Z11}, 
but Ref.\ \cite{Z12} reports equi\-valent results only for motion 
through a crosslinked gel, where constraint release is suppressed. 
The corresponding melt shows quite different behavior.

Invoking tube length fluctuations and constraint release we qualitatively 
may interpret the observed deviations from the asymptotic reptation 
results as crossover behavior outside the region of asymptotic chain 
lengths. However, there exist other theories of melt dynamics, which 
are not based on the tube concept and describe many experiments as well 
\cite{Z13,Z14}. (See also the review \cite{Z7}.) It thus is conceivable 
that the basic assumptions of the reptation scenario do not hold. 
To get more insight into these problems, we clearly have to quantitatively 
evaluate the consequences of the pure reptation model, beyond asymptotic 
power laws. 

In previous analytical work \cite{Z15,Z16} we determined the motion of 
individual segments of the chain. Since our theory involves some 
approximations, we compared the results to simulations \cite{Z17} 
of the Evans-Edwards model \cite{Z18}, which is an accurate implementation 
of the pure reptation model. In essence, both theory and simulations 
agreed in showing that the crossover among various asymptotic power 
law regions is very slow. The crossover regions are so broad that 
the asymptotic power laws can be identified only for very long chains. 
For example, using the Evans-Edwards model with the smallest possible 
tube diameter we could identify the $t^{1/4}$-law for the motion of 
the central segment only for chain lengths $N \agt 160$. This law is 
the easiest to observe, and our evaluation of the theory predicts that 
other asymptotic laws unambiguously can be identified only for much 
longer chains. This result is in line with the known slow crossover 
behavior of the reptation time \cite{Z6}, which is predicted to reach 
the asymptotic law $T_3 \sim N^3$ only for chain lengths far beyond 
present day experimental feasibilities. Still, for the motion of 
individual segments, the full crossover functions can be calculated, 
and our analytical results very well agree with our simulations. 
Furthermore, for shorter chains all our analytical and simulational 
results qualitatively are very similar to results of simulations of 
melts \cite{Z9,Z10}. In later work \cite{Z19} we considered chain 
motion in a time independent, but disordered environment, where 
the disorder affects only the chain mobility but leaves 
the equilibrium distribution of chain configurations unchanged. 
We found that with such 'kinematic' disorder reptation prevails, 
an observation which recently has been supported \cite{Z20} by 
rigorous bounds on the diffusion coefficient. Since such kinematic 
disorder certainly is present in a melt, these results again support 
reptation as adequate theory of melt dynamics. 
   
In contrast to the motion of individual segments the coherent structure 
function $S_{c}(q,t;N), \: (q = |{\bf q}|$ : scattering vector, $t$ : time), 
is well accessible in real experiments. It can be measured, for instance, 
by neutron scattering from a mixture of deuterated and hydrogenated chains. 
As mentioned above, approximate asymptotic forms for $S_{c}(q,t;N)$, 
based on reptation type theories, can be found in the literature \cite{Z4,Z5}. 
However, as for the motion of individual segments we can evaluate the full 
reptation theory for $S_{c}(q,t;N)$ also outside the asymptotic regime. 
As for individual segments, we then expect to see important preasymptotic 
or crossover effects. The evaluation of $S_{c}(q,t;N)$ including full 
reptational dynamics is somewhat complicated, and our theory in detail 
has been presented in Ref. \cite{Z3}. Here we compare the results to 
simulations of the Evans-Edwards model. We consider both the scattering 
from the total chain and from interior subchains. The latter is important 
to estimate the reliability of the theory, which can treat end-effects 
only in some approximation.

In Section 2 we discuss our simulations. Section 3 is devoted to a 
comparison with the expressions for $S_{c}(q,t;N)$ given by Doi and 
Edwards \cite{Z4,Z2} or by de Gennes \cite{Z5}, respectively. 
In Section 4 we qualitatively describe our theory and give an empirical 
fit formula which reasonably well describes our quantitative results. 
The comparison between our theory and our Monte Carlo data is presented 
in Sect.\ 5. Section 6 contains our conclusions.
  
\section{Simulations}\label{kap2}

\subsection{The Evans-Edwards model}

A very simple model for simulating reptation has been introduced by 
Evans and Edwards \cite{Z18}. The configuration 
$\{{\bf r}_{1},\ldots,{\bf r}_{N}\}$ of the chain is taken as 
a random walk of $N - 1$ steps $|{\bf r}_{j} - {\bf r}_{j-1}| = \ell_{0}$ 
on a cubic lattice. The lattice spacing $\ell_{0}$ henceforth defines 
the unit of length. The obstacles are taken as the edges of the dual 
lattice. In the interior of the chain, the obstacles suppress any motion 
except for the motion of `hairpins', i.e., configurations of three 
subsequent beads of type 
$\{{\bf r}_{j-1},{\bf r}_{j},{\bf r}_{j+1} = {\bf r}_{j-1}\}$. 
In an elementary move the tip ${\bf r}_{j}$ of the hairpin with 
equal probability hops to any of the six neighbors of the site 
${\bf r}_{j-1} = {\bf r}_{j+1}$. The chain ends ${\bf r}_{1},{\bf r}_{N}$ 
are free to hop between all neighbors of ${\bf r}_{2},{\bf r}_{N - 1}$, 
respectively. Fig.~1 shows a sequence of internal configurations resulting 
from this dynamics. In our simulations we used the same implementation of 
the model as in our previous work \cite{Z17}, and we measured the coherent 
structure function defined as 
\begin{eqnarray}
S_{c}(q,t^{(MC)};N) 
&=& \sum_{j_{1},j_{2}=1}^{N} 
\left\langle \overline{e^{i {\bf q} 
\left({\bf r}_{j_{1}}(t^{(MC)}) - {\bf r}_{j_{2}}^{(0)}\right)}}\right\rangle
\nonumber \\
&=& \left\langle \overline{{\cal C} ({\bf q},t^{(MC)}) {\cal C} ({\bf q},0) 
+ {\cal S} ({\bf q},t^{(MC)}) {\cal S} ({\bf q},0)} \right\rangle
\nonumber\\
\end{eqnarray}
where 
\begin{eqnarray}
{\cal C}({\bf q},t) &=& \sum_{j = 1}^{N} \cos ({\bf q} {\bf r}_{j}(t))
\nonumber \\
{\cal S}({\bf q},t) &=& \sum_{j = 1}^{N} \sin ({\bf q} {\bf r}_{j}(t))~.
\end{eqnarray}
(The imaginary part ${\cal S}({\bf q},t) {\cal C}({\bf q},0) - 
{\cal C}({\bf q},t) {\cal S}({\bf q},0)$ of $S_{c}$ is zero on average, 
of course.) To get more information on the internal motion, we also 
measured the coherent structure function $S_{c}(q,t^{(MC)};M,N)$ of 
the central subchain of length $M$, defined by restricting the summations 
in Eqs.\ (2.1), (2.2) to the interval $[(N - M)/2 + 1, (N + M)/2]$. 
>From our previous work, we expect to see features characteristic of 
reptation for $N \agt 100$, and we therefore used chain lengths 
$N = 80,\;160,\;320$. Since for our model, the entanglement length 
is estimated as $N_{e} \approx 3.7$ (see Sect.\ 4.B), this yields values 
$22 \alt N/N_{e} \alt 87$. Similar values are extracted from many 
simulations or experiments on melts, so that our results should be 
relevant also for the interpretation of such data.
Monte Carlo time $t^{(MC)}$ is measured in units of one attempted move 
per bead on average. We performed runs up to 
$t^{(MC)}_{\rm max} = 5 \cdot 10^{10}$, and we measured the structure 
function up to $t^{(MC)} = 4.5 \cdot 10^{9}$ using a moving time average. 

\subsection{Statistical accuracy}

During a run, the longer chains do not diffuse very far, and data of 
a single run therefore are strongly correlated. To get reasonably 
accurate results we have to average over many independent runs. A priori, 
this poses a problem for larger momenta, $q R_{g} \agt 1$, where $R_{g}$ 
is the radius of gyration of the chain. It is easily checked that the 
reduced width of the distribution of the static structure function 
with increasing $q R_{g}$ rapidly tends to $1$. Indeed, in the limit 
$N \rightarrow \infty, q > 0$ fixed, the distribution of $S_{c}(q,0;N)$ 
approaches the exponential distribution. This suggests that $10^{4}$ 
runs are needed to reduce the statistical uncertainty to a few percent. This would pose no problems, if we just were interested in static properties. However, for the longer chains a single run extending to times well beyond the reptation time takes several hours on a standard work station. In measuring {\em dynamic} quantities the number of runs therefore inevitably is much smaller than needed for a precise determination of {\em static} quantities.

Fortunately it turns out that the dynamics essentially is decoupled from the static configuration. This is illustrated in Fig.~2, which shows results for 
$S_{c}(q,t^{(MC)};N), q \ell_{0} = 0.5, N = 320$, normalized to the 
exact static structure function $S_{0}(q,N)$ of the model, which easily 
is calculated analytically (see Sect.\ 2.D).  Each curve in Fig.~2~a 
is averaged over $10^{3}$ independent short runs $(t^{(MC)} \leq 10^{5})$, 
including the moving time average for each run. Clearly the scatter of 
$S_{c}(q,0;N)$ is consistent with the above discussion. It is larger 
than the temporal variation of the curves. However, plotting in Fig.~2~b 
the normalized time dependence 
$[S_{c}(q,t^{(MC)};N) - S_{c}(q,0;N)]/S_{0}(q,N)$, we find 
that all curves nearly coincide. The global dynamics measured 
by the structure function, is not correlated with the static 
configuration, an observation which supports one of the basic 
assumptions of the reptation model. The reason behind that observation 
is easily understood. The mobility of the chain is governed by the number 
of hairpins which essentially is Gaussian distributed and fluctuates much 
less than the static structure function. Furthermore, except for rare 
events, viz. extremely stretched or extremely compact configurations, 
the number of hairpins is independent of the overall (tube-) conformation 
of the chain. With this insight, we take as basic data the difference 
$S_{c}(q,t^{(MC)};N) - S_{c}(q,0;N)$ for each run. The error bars in 
our plots give twice the standard deviation of this difference. 
We always plot the normalized dynamic structure function defined as 
\begin{eqnarray}
\lefteqn{\bar{S}_{c}(q,t^{(MC)};N)= 1 + \frac{1}{S_{0}(q,N)}\;\cdot}\\
&& \quad\Big(S_{c} 
\left(q,t^{(MC)};N\right) - S_{c} (q,0;N)\Big)_{\mbox{averaged~over~runs}}~,
\nonumber
\end{eqnarray}
where $S_{0}(q,N)$ is the exact static structure function, not the 
measured average value of $S_{c}(q,0;N)$. Depending on the number of 
independent runs, these two quantities differ by $0.1 - 6 \%$. 

For reasons of computer memory, we performed runs with three different 
values of the maximal time $t^{(MC)}_{\rm max}$. 
To observe the small initial effects we for each value of chain length 
$N$ and wave vector $|{\bf q}|$ averaged over $10^{4}$ short runs with 
$t^{(MC)}_{\rm max} = 10^{5}$, taking data up to 
$t^{(MC)} = 1.4 \cdot 10^{4}$. These data are denoted by small dots 
in the later figures. In the intermediate time range, 
$t^{(MC)}_{\rm max} = 10^{8}$, we measured the structure function 
for $10^{4} \leq t^{(MC)} \leq 10^{7}$ and always performed 50 runs 
(big dots in the figures). In the long time range,
$t^{(MC)}_{\rm max} = 5 \cdot 10^{10}$, we took data for 
$10^{5} \leq t^{(MC)} \leq 4.5 \cdot 10^{9}$, averaging over 
30 to 100 runs (circles in the figures). For each set of runs with 
given maximal time, we also measured the average of $S_{c}(q,0;N)$.   

For additional information on the statistical accuracy of our data, 
we measured the imaginary part of $S_{c}(q,t^{(MC)};N)$, which 
rigorously vanishes for $t^{(MC)} = 0$, but fluctuates about zero 
for $t^{(MC)} > 0$ in a finite sample. For longer times we typically 
found average values of order $\pm 0.01 S_{0}(q,N)$, again smaller 
than the uncertainty of $S_{c}(q,0;N)$ in long-time runs. 

\subsection{Momentum range}

First considerations suggest to restrict the analysis to the momentum 
range $R_{g}^{-2} \ll q^{2} \ll \ell_{0}^{-2}$. Momenta of order 
$q^{2} R_{g}^{2} \alt 1$ do not resolve the tube but rather see a 
cloud of beads. For momenta $q^{2} \ell_{0}^{2} \agt 1$ the 
micro-structure of the chain might play a role. In our simulations 
the above condition can only poorly be satisfied. For our longest 
chain $(N = 320)$ it reads $0.02 \ll q^{2} \ell_{0}^{2} \ll 1$, 
leaving at best a small window close to $q \ell_{0} \approx 0.4$.

However, a closer inspection reveals that these considerations do not 
seriously restrict dynamic measurements. To check the relevance of the 
condition $q^{2} \ell_{0}^{2} \ll 1$ we analyzed the dynamics of a free 
chain. As is well known, asymptotically the standard Rouse dynamics is 
found if for a lattice chain  `kink jumps': 
$\{{\bf r}_{j} - {\bf r}_{j-1} = {\bf s}_{1}, \;
{\bf r}_{j+1} - {\bf r}_{j} = {\bf s}_{2}\} \Rightarrow 
\{{\bf r}_{j} - {\bf r}_{j-1} = {\bf s}_{2},\; 
{\bf r}_{j+1} - {\bf r}_{j} = {\bf s}_{1}\},~
{\bf s}_{1} \cdot {\bf s}_{2} = 0 $ are allowed besides hairpin-moves. 
We found that with this modified dynamics, the normalized coherent 
structure function is in excellent agreement with the relaxation function 
calculated from the continuous chain Rouse model, even for $q \ell_{0} = 1$. 
Micro-structure effects arising from the finite segment size die out 
very rapidly on the scale of about 10 $MC$-steps and therefore should 
be negligible also for reptation dynamics. Of course this does not exclude 
the possibility of dynamical micro-structure effects which might influence 
the short time regime and are not accounted for by the reptation model. 
In particular, reptation does not properly treat the dynamics of 
fluctuations perpendicular to the tube axis.

Concerning the condition $q^{2} R_{g}^{2} \gg 1$, we note that smaller 
wave vectors, of course, provide little information on the details of 
the internal motion of the chain, but at least they measure the global 
motion of the coil. However, the slowness of this motion asks for extremely 
large time ranges $t^{(MC)}_{\rm max} \gg T_{3}$. 
We therefore performed only one series of simulations for 
$|{\bf q}| \ell_{0} = 0.1,\; N = 160$, corresponding to 
$q^{2} R_{g}^{2} = 0.267$. Most of our simulations use values 
$3 \alt q^{2} R_{g}^{2} \alt 50$, with a maximal value of 
$q^{2} R_{g}^{2} \approx 53$ reached for $|{\bf q}| \ell_{0} = 1, N = 320$. 

\subsection{Normalization}

In our simulations we choose ${\bf q}$ to point into one of the three lattice directions. 
For that choice, the static correlations between segments $j$  and $k$ 
take the form
\begin{eqnarray}
\left\langle e^{i {\bf q} ({\bf r}_{j}(0) - {\bf r}_{k}(0))}\right\rangle 
&=& \left(\frac{2}{3} + \frac{1}{3}\:\cos\:|{\bf q}|\right)^{|k-j|}
\nonumber \\
&=& \exp (- \bar{q}^{2} |k - j|) \: \: \:,
\end{eqnarray}
where
\begin{equation}
\bar{q}^{2} = - \ln \left(\frac{2}{3} + \frac{1}{3}\:\cos\:|{\bf q}|\right)~.
\end{equation}
Recall that the lattice constant $\ell_{0}$ defines the unit of length.
Summing the segment indices over the chain, we find the static structure 
function which is used in normalizing our results:
\begin{eqnarray}
S_{0} (q,N) &=& \sum^{N}_{j,k = 1}\:\exp (- \bar{q}^{2} |k-j|)
 \\
&=& N + \frac{2}{(e^{\bar{q}^{2}} - 1)^{2}} 
\left[e^{- \bar{q}^{2} (N - 1)} - N + (N - 1)\; e^{\bar{q}^{2}} \right]~.
\nonumber
\end{eqnarray}
In our simulations, we for each run averaged over all three lattice directions.

\section{Comparison of data and simplified reptation type theories}\label{kap3}

\subsection{Asymptotic form of $S_{c}(q,t;N)$ derived by Doi and Edwards}

A simplified version of the reptation model concentrates on the dynamics 
of the `primitive chain' \cite{Z2,Z4}, which is a reduced form of the chain, 
lying stretched in the tube. In the Evans-Edwards model the primitive chain 
can be viewed as the non-reversal random walk derived from the random walk 
configuration of the physical chain by cutting off all hairpins. 
All internal degrees of freedom are neglected so that within time 
interval $\Delta t$, all parts of the primitive chain move the same 
distance $\Delta s$ along the tube. The length of the primitive chain 
is taken to be fixed. (This model often is addressed as 
`the reptation model'. We will use the term `primitive chain model' to 
distinguish it from the full reptation model which deals with the 
dynamics of spared length as the elementary process.)

With its simplifications, the primitive chain model treats only 
the destruction of the initial tube, as resulting from the global 
motion of the chain. It neglects tube length fluctuations which are 
due to the uncorrelated motion of the chain ends as well as relaxation 
in the interior of the tube. Since both these additional effects are 
governed by the equilibration time $T_{2}$ of the chain, the primitive 
chain model is restricted to times $t \gg T_{2}$. Furthermore, 
the segment indices are taken as continuous variables, and this 
`continuous chain limit' restricts the theory to long chains: $N \gg 1$. 

Within this model the time dependent correlation function of two beads:
\begin{equation}
S (q,t;j,k,N) = \left\langle \overline{\exp \left[ i {\bf q} ({\bf r}_{j}(t) - {\bf r}_{k}(0))\right]}\right\rangle\: \: \:,
\end{equation}
obeys a diffusion equation. Solving this equation and integrating $j,\;k$ 
over the chain, Doi and Edwards arrive at the following result for 
the normalized coherent structure function 
(see Ref.~\cite{Z2}, chapter 6.3.):
\begin{equation}
\bar{S}_{c} (q,t;N) = \frac{S_{c}(q,t;N)}{S_{c}(q,0;N)} 
= \bar{S}_{DE} (q^{2} R_{g}^{2}, t/T_{3})
\end{equation}
\begin{equation}
\bar{S}_{DE} (Q,\tau) = \frac{1}{D(Q)} \sum^{\infty}_{p=1} 
\frac{Q \sin^{2} \alpha_{p}}{\alpha_{p}^{2} (Q^{2}/4 + Q/2 
+ \alpha_{p}^{2})} \exp \left( - \alpha_{p}^{2} \tau \right) \: \: \:.
\end{equation}
Here the $\alpha_{p}$ are the positive solutions of 
\begin{equation}
\alpha_{p} \tan \alpha_{p} = \frac{1}{2} Q = \frac{1}{2} q^{2} R_{g}^{2}~,
\end{equation}
and
\begin{equation}
D (Q) = \frac{2}{Q^{2}} (e^{-Q} - 1 + Q)
\end{equation}
is the Debye function. With the assumptions of the theory, the reptation 
time attains its asymptotic behavior $T_{3} \sim N^{3}$. Note that 
$T_{3}$ is related to the time scale $\tau_{d}$ introduced in 
Ref.~\cite{Z2} by $T_{3} =  \frac{\pi^{2}}{4} \tau_{d}$.

The result (3.2),(3.3) for $\bar{S}_{c} (q,t;N)$ depends only on 
$Q = q^{2} R_{g}^{2} \sim \bar{q}^{2} N$ and $\tau = t/T_{3} \sim t/N^{3}$, 
but not on $N$ separately. To test this feature, we carried through 
simulations for $N = 320, q = 0.25$ and $N = 80, q = 0.5024$, both 
parameter sets resulting in $q^{2} R_{g}^{2} \approx 3.3$. Fig.~3 
compares our simulation results to $\bar{S}_{DE}$, plotted as function 
of $\log_{10}(t/T_{3})$. The Monte Carlo time has been scaled so that 
in the region $t/T_{3} \agt 1$, the data for $N = 320$ fit to the 
theoretical curve. In view of $T_{3}\sim N^{3}$, for $N = 80$ an 
additional factor $4^{3}$ has been included in the time scale. 
With this scaling, the deviation between the two sets of data shown 
in Fig.~3 proves that we have not yet reached chain lengths large 
enough for the primitive chain model to hold.

Still, for larger times, $\bar{S}_{DE}$ and the data for $N = 320$ 
agree quite well. This suggests to treat $T_{3} = T_{3} (N)$ as a 
fit parameter in adjusting the data to 
$\bar{S}_{DE} (q^{2} R_{g}^{2}, t/T_{3})$, giving up the strict 
proportionality $T_{3} \sim N^{3}$. However, a closer inspection 
of Fig.~3 reveals that the data initially decrease faster than 
$\bar{S}_{DE}$. This is a systematic effect, observed for all chain 
lengths and wave vectors. It points to the influence of relaxation modes 
neglected in the primitive chain model. According to a suggestion of 
de Gennes \cite{Z5}, internal relaxation, in particular, yields an 
initial decrease of $\bar{S}_{c} (q,t;N)$ which saturates at some 
$q$-dependent plateau value. We thus should fit the data to 
$B_{DE} \bar{S}_{DE} (q^{2} R_{g}^{2}, t/T_{3})$, with $B_{DE}$ 
as another free parameter.

A detailed analysis of the reptation model points to tube length 
fluctuations as the origin of the initial decrease, rather than 
internal relaxation. (See Sect.\ 4.C and Ref.~\cite{Z3}.) Still we 
may fit our data to $B_{DE} \bar{S}_{DE}$, where $B_{DE} = B_{DE}(q,N)$, 
$T_{3} = T_{3} (N)$. Fig.~4 shows the results for $N = 320$. 
We clearly find very good agreement among theory and data in the 
time region $t \agt 20 T_{2}$. (The estimate for $T_{2}$ has been 
taken from our theory, see Sect.\ 4.B.) Deviations occur for smaller times, 
which is consistent with the approximations inherent in the primitive 
chain model. Within the realm of that model the central segment moves 
according to $\left\langle \overline{(r_{N/2}(t) - r_{N/2}(0))^{2}}
\right\rangle \sim (t/N)^{1/2}, T_{2} \ll t \ll T_{3}$ (cf.\ Eq.\ (1.1)), 
and from our previous work \cite{Z17} we know that this law certainly is 
not attained before $t \agt 20 T_{2}$. The results shown in Fig.~4 are 
typical also for other chain lengths.

Since $T_{3}$ now plays the role of an effective fit parameter, 
it a priori could depend both on $N$ and $q$. Parameters 
$T_{3} = T_{3} (q,N)$, $B_{DE} = B_{DE} (q,N)$ extracted by fitting 
our data are collected in Table 1. Any $q$-dependence of $T_{3}$ is 
found to be weak, if significant at all. $B_{DE}$ depends on $q$, but is essentially 
independent of $N$. 

Ignoring any $q$-dependence, we in Fig.~5 have plotted the values 
of $T_{3}(N)$ normalized to the asymptotic behavior $T^{\infty}_3$ 
resulting from Eqs.\ (4.5), (4.7), (4.12) below. It must be stressed 
that we here {\em define} a reptation time in terms of the large time 
behavior of the scattering function (and the Doi-Edwards form 
$\bar{S}_{DE}$). Other definitions based on other observables 
(or other theoretical expressions) may yield somewhat different results. 
In previous work \cite{Z17}, we defined a reptation time $\tilde{T}_{3}$ 
by the criterion that the mean squared distance moved by a chain end 
equals the equilibrium mean squared end-to-end distance: 
$\left\langle \overline{({\bf r}_{0}(\tilde{T}_{3}) - 
{\bf r}_{0}(0))^{2}}\right\rangle = R_{e}^{2}$. The previous data 
are included in Fig.~5 (open circles). The two definitions yield 
somewhat different results for the corrections to the asymptotic 
behavior. In the range of $N$ measured here, $\tilde{T}_3$ is about 30 \% below $T_3$ as taken from the scattering function. It, however, must be noted that the values extracted from 
the scattering functions depend somewhat on the time range included 
in the fit. We estimate this uncertainty to be of the order of 5 \%. 
In Fig.~5 we included lines corresponding to effective power laws 
$T_3 \sim N^z$. The effective powers are consistent with expectations 
based on previous work \cite{Z6}.

To summarize, our results show that the coherent structure function 
of the primitive chain model may well be used to fit data for large 
times, $t \agt 20 T_{2}$, provided we allow for some phenomenological 
prefactor $B_{DE}$ and take the reptation time $T_{3}$ as an effective 
parameter defined by the fit. For smaller times, deviations are seen, 
that increase with increasing $q$. 

\subsection{Comparison to Gennes' theory}

In his work \cite{Z5}, de Gennes considers only intermediate values 
of $q$: $\ell_{0}^{-2} \gg q^{2} \gg R_{g}^{-2}$ and constructs the 
scattering function as a sum of two terms. For $t \gg T_{2}$ the 
`creep' term dominates. This term is just the limiting form of 
$\bar{S}_{DE}$ (Eq.\ (3.3)) for $Q \rightarrow \infty$:
\begin{equation}
\bar{S}^{(c)} (t,N) = \frac{8}{\pi^{2}} \sum^{\infty}_{p=0} 
(2 p + 1)^{-2} \exp \left[ - (2 p + 1)^{2} \frac{\pi^{2}}{4} 
\frac{t}{T_{3}} \right] \: \: \:.
\end{equation}
It tends to $1$ for $t/T_{3} \rightarrow 0$. It is combined with 
some contribution of local relaxation, which is calculated in two steps. 
First the interior relaxation of a Rouse chain in one dimensional space 
is calculated, where the chain is stretched so that the end-to-end 
distance equals the tube length. In the second step this one dimensional 
motion is embedded into the three dimensional random walk configuration 
of the tube. Tube length fluctuations are neglected. With some additional 
approximation, this model of a Rouse chain in a coiled tube yields a 
`local' contribution
\begin{equation}
\bar{S}^{(\ell)} (q,t) = e^{t_{1}}\:\mbox{erfc}\: \sqrt{t_{1}} \: \: \:,
\end{equation}
where 
\begin{equation}
t_{1} = \frac{3}{4} \frac{N}{N_{e}} (q^{2} R_{g}^{2})^{2} \frac{t}{T_{3}} \: \: \:.
\end{equation}
Since again the validity of the asymptotic law $T_{3} \sim N^{3}/N_{e}$ 
is assumed, the chain length $N$ and the entanglement length $N_{e}$ drop 
out in Eq.\ (3.8), as expected for a contribution resulting from strict 
one dimensional internal relaxation. $\bar{S}^{(\ell)} (q,t)$ describes 
internal relaxation on length and time scales exclusively determined by $q$. 
The total result for the normalized coherent scattering function reads
\begin{eqnarray}
\lefteqn{\bar{S}_{c} (q,t;N) = \bar{S}_{dG} (q,t;N) =}\\
&&\quad 
(1 - B_{dG}) \bar{S}^{(\ell)}(q,t) + B_{dG} \bar{S}^{(c)}(t,N) \: \: \:,
\nonumber
\end{eqnarray}
where according to de Gennes
\begin{equation}
B_{dG} = B_{dG}(q) = 1 - \frac{N_{e}}{36}\:q^{2} \ell_{0}^{2}
\end{equation}
is independent of $N$.

In an attempt to extend the range of wave vectors to $q \ell_{0} \agt 1$, 
Schleger et al \cite{Z21} used $B_{dG}(q) 
= \exp (- q^{2} \ell_{0}^{2} N_{e}/36)$. With this choice it was 
found \cite{Z21,Z22,Z23} that $\bar{S}_{dG} (q,t;N)$ over a large 
range of wave vectors and for several chain lengths yields a good fit 
to neutron scattering data from melts of Polyethylene. However, 
P\"utz et al \cite{Z10}, using the same form of $\bar{S}_{dG} (q,t;N)$ 
to analyze simulation data for melts of very long chains, found only 
poor agreement. In later work \cite{Z24} they argued that this was due 
to some ambiguity in the relation between entanglement length $N_{e}$ 
and tube diameter. 

In our recent work \cite{Z3} we reconsidered the model of a Rouse chain 
in a tube. Our analysis reveals some serious deficiency of this model: 
it does not start from equilibrium initial conditions. An ensemble of 
stretched one dimensional random walk chains folded into the three 
dimensional random walk configuration of the tube is not identical 
to the equilibrium ensemble of three dimensional random walk chains. 
The local structure of the chains differs on the scale of the tube 
diameter. For the static structure function, this yields a correction 
of relative order $N_{e}/N$ that vanishes in the limit $q^{2} R_{g}^{2} 
= \mbox{const}, N \rightarrow \infty$. For the time dependence, however, 
the effect is serious since the non-equilibrium initial conditions relax 
only on scale $T_{2}$. Our analysis shows that this unphysical relaxation 
indeed dominates the time dependence of the `local' contribution 
$\bar{S}^{(\ell)}(q,t)$, as calculated from this model (see Fig.~3 
of Ref.~\cite{Z3}). Data analysis based on Eq.\ (3.9) with 
$\bar{S}^{(\ell)}(q,t)$ taken from the model of a Rouse chain 
in a tube therefore is not particularly meaningful. 

Still, in view of the experimental findings cited above, we may ask 
whether Eqs.\ (3.6) - (3.9) can be used as heuristic modeling of 
the coherent structure function resulting from reptation. Fig.~6 
shows the result for the longest chain $(N = 320)$ and largest 
wave-vector $(q = 1.0)$ measured. The resulting value 
$q^{2} R_{g}^{2} \approx 53.3$ is large enough for a reasonable 
test of the form (3.9), which assumes $q^{2} R_{g}^{2} \gg 1$. 
Replacing in Eq.\ (3.9) $\bar{S}_{c}$ by the full result $\bar{S}_{DE}$ 
of the primitive chain model does not seriously change the picture. 
The reptation time is fixed by the long-time tail and is taken from 
Table 1. For $B_{dG}$, we used the form suggested by Schleger et al: 
$B_{dG} = \exp (- q^{2} \ell_{0}^{2} N_{e}/36)$. $N_{e}$ in principle 
is known from our previous analysis of segment motion: $N_{e} = 3.69$ 
(see Sect.\ 4.B). Thus all parameters are fixed, and the result for 
$\bar{S}_{dG}$ (full curve in Fig.~6) strongly deviates from the data, 
except for the extreme long-time tail. This is no surprise since the value $B_{dG} \approx 0.9$, resulting for $N_{e} = 3.69$, considerably exceeds the value $B_{DE} \approx 0.78$ extracted from fitting with the Doi-Edwards form. (See Table 1.) If we treat $B_{dG}$ as a free parameter, we clearly in the range $t \agt 20 T_2$ can enforce agreement among theory and data, (see Fig.~4), at the expense of considerably underestimating $\bar{S}_{c}$ for $t \alt T_2$. The situation can be improved only, if we also change the scale of $t_{1}$, dividing $t_1$, (Eq.~3.8), by a factor of order 200. Quite similar 
results are found for $N = 160$, where $t_{1}$ has to be divided by 
a factor of order 50 to reproduce the average trend of the data for 
$t \alt T_{2}$.

We conclude that also from a purely phenomenological point of view, 
the form (3.9) of $\bar{S}_{c}$ is not justified. The large and chain 
length dependent rescaling required for $t_{1}$ suggests that tube 
length fluctuations governed by time scale $T_{2}$ might be much more 
important than internal relaxation processes governed by an 
$N$-independent scale.         

We finally note that in Ref.\ \cite{Z23} a modified form of Eq.\ (3.9) 
has been proposed, in which the creep term $\bar{S}^{(c)}$ is calculated 
from tube length fluctuations, but the form (3.7) of the local contribution 
is retained. The analysis is restricted to the time range $t \ll T_2$, 
which is the relevant range for neutron scattering experiments. 
From Fig.~6 it is clear that with the proper parameter values 
$N_e = 3.69, \: t_1$ as given in Eq.\ (3.8), this modification cannot 
improve the fit for $t \ll T_2$, since tube length fluctuations only 
decrease the contribution $\bar{S}^{(c)}$. Again a reasonable fit can be reached only if we take both $B_{dG}$ and the scale of $t_1$ as free parameters.

\section{Results of the full reptation model}

\subsection{Basic ideas of our approach}

The primitive chain model neglects all internal degrees of freedom, 
and an attempt to model internal relaxation as one dimensional Rouse 
motion leads to unphysical initial conditions, which seriously affect 
the results up to times of the order of the internal relaxation time 
$T_{2}$. However, the simplifying assumptions underlying these approaches are 
no essential part of the original reptation model. As recalled in the 
introduction, reptation as single elementary process involves the 
diffusion of spared length along the chain, together with its decay 
and creation at the chain ends. The separation of the dynamics into 
internal relaxation, tube length fluctuations and motion of the 
primitive chain therefore is somewhat artificial. In particular, 
tube length fluctuations cannot properly be separated from the motion 
of the primitive chain, as will be discussed in Sect.\ 4.C. Furthermore, 
from our analysis of segment motion we know that for typical experimental 
chain lengths we are in a crossover region where all dynamic effects must 
be treated on the same level by evaluating the consequences of full 
reptational dynamics. 

In our analytical work \cite{Z16,Z3} we use a very simple 
implementation of the reptation idea, in which the wiggles 
of spared length are represented by particles which hop along 
the chain, with hopping probability $p$ per time step. The particles 
do not interact, and a given particle sees the others just as part 
of the chain. If a particle passes a bead, it shifts the position 
of this bead in the tube by the spared length $\ell_{S}$. The chain 
ends are coupled to large particle reservoirs, which absorb and emit 
particles at such a rate that the equilibrium density $\rho_{0}$ of 
spared length on the chain is maintained on the average. Absorption or 
emission of a particle prolongs or shortens the tube at the corresponding 
chain end by the spared length $\ell_{S}$. Keeping track of the change 
in the occupation number of the reservoirs, we therefore control the 
tube length fluctuations as well as the destruction of the original tube. 
In particular, within time interval $[0,t]$ an end piece of length 
$\ell_{S} n_{\rm max}(t)$ of the original tube has been destroyed, 
where $n_{\rm max}(t)$ is the largest negative fluctuation of the 
occupation number of the corresponding reservoir during this time interval.

All moments of the stochastic processes which determine the motion 
of internal segments or the occupation of the reservoirs, can be 
evaluated rigorously, but the determination of the maximal fluctuation 
$n_{\rm max}(t)$ poses a serious problem. The occupation number of a 
reservoir carries out a {\em correlated} random process, since a particle 
emitted can be reabsorbed by the same reservoir later. This correlation 
dies out only on scale of the internal equilibrium time $T_{2}$ of 
the chain. It does not prevent the evaluation of arbitrary moments, 
but the maximal fluctuation cannot be calculated rigorously.  
Such a calculation is possible \cite{Z25} only for an {\em uncorrelated} 
random process. To determine the degree of tube destruction, we therefore 
have to resort to some approximation. In our method, we basically for each 
final time $t$ replace the correlated random process by that uncorrelated 
random walk which for this time yields the correct moments. The effective 
hopping rate of this random walk depends on the final time $t$. It changes 
from the microscopic rate $\rho_{0} \cdot p$ for $t \ll T_{2}$ to the 
mobility $\rho_{0} \cdot p/N$ of the primitive chain for $t \gg T_{2}$, 
which is a physically most reasonable behavior.

In essence, this `mean hopping rate' approximation for the coherent 
structure function smoothly interpolates between two rigorously 
accessible limits. For $t \ll T_{2}$, tube renewal does not influence 
the motion  of an interior piece of the chain, and the coherent structure 
function of such pieces can be calculated rigorously. For $t \gg T_{2}$ 
the correlations of the stochastic processes are negligible and 
the mean hopping rate approximation should become exact. Indeed, 
we find that in this limit the motion of all chain segments is 
tightly bound to the motion of the chain ends. As a result, 
the problem reduces to the uncorrelated motion of a single 
stochastic variable, as in the primitive chain model. 
The details of our approximation are discussed extensively in 
Refs.\ \cite{Z3,Z16}, and will not be repeated here. 

The coherent scattering function can be determined by summing 
the contribution of two beads
\begin{equation}
S (q,t;j,k,N) = \left\langle \overline{e^{i {\bf q} 
({\bf r}_{j}(t) - {\bf r}_{k}(0))}}\right\rangle
\end{equation}
over the bead indices $j,k$. For this function we in Ref.~\cite{Z3}, 
Sect.\ 6.A, have derived an integral equation of the form 
\end{multicols}
\begin{eqnarray}
S (q,t;j,k,N) &=& S^{(T)}(q,t;j,k,N)
\nonumber \\ 
&+& \int_{0}^{t}\:d t_{0} \sum^{N}_{j_{0} = 0} 
\Big\{{\cal P}^{*}(j_{0},t_{0}|0) \exp ( - \bar{q}^{2} |j_{0} - k|) 
S (q, t-t_{0};j,0,N) 
\nonumber \\
&+& {\cal P}^{*} (j_{0},t_{0}|N) \exp ( - \bar{q}^{2} |j_{0} - k|) 
S (q, t-t_{0};j,N,N)\Big\}\: \: \:.
\end{eqnarray}
\begin{multicols}{2}
Here $d t_{0} {\cal P}^{*}(j_{0},t_{0}|m), m = 0,N$, is the probability 
that the initial tube is finally destroyed within time interval 
$[t_{0},t_{0} + d t_{0}]$, its last piece being the initial position 
of segment $j_{0}$, which at time $t_{0}$ is occupied by chain end $0$ 
or $N$, respectively. The exponential factors result from the random walk 
configuration of the tube, cf.\ Eqs.\ (2.4), (2.5). The inhomogeneity 
$S^{(T)}(q,t;j,k,N)$ is the contribution to $S (q,t;j,k,N)$ of all those 
stochastic processes, which do not completely destroy the original tube.

Summing Eq.\ (4.2) over the segments $j,k$ we find a system of 
two equations which have to be solved numerically. The kernel 
${\cal P}^{*}$ and the inhomogeneity $S^{(T)}$ are determined 
within the mean hopping rate approximation. The results are lengthy 
and will not be reproduced here. We rather in Subsect.\ D give an 
analytical expression, which in the range of wave vectors and chain 
lengths considered in the present work, reasonably well reproduces 
the numerical results of our theory.

We finally note that we analytically can prove (Ref.~\cite{Z3}, Sect.\ 7) 
that our theory in the limit of infinite chain length $N \rightarrow \infty$, 
with $t/T_{3}$ and $q^{2} R_{g}^{2}$ kept fixed, reproduces the result 
of the primitive chain model. Also the relation to the model of a Rouse 
chain in a tube can be analyzed in precise terms, if we consider 
an interior piece of length $M$ in an infinitely long chain. 
We find (Ref.~\cite{Z3}, Sect.\ 5) that this model reproduces the 
results of reptation only for $t/T_{2}(M) \rightarrow \infty$, 
$q^{2} R_{g}^{2}(M)$ fixed, where $T_{2}(M)$ or $R_{g}(M)$ are 
the equilibration time and the radius of gyration of the subchain 
considered. For $t/T_{2}(M) \alt 1$ the non-equilibrium initial 
condition seriously affect the scattering function, as has been discussed in Sect.\ 3.B.

\subsection{Microscopic parameters of the reptation model}

The microscopic parameters of the model are the segment size 
$\ell_{0} = |{\bf r}_{j} - {\bf r}_{j-1}|$, the average density 
$\rho_{0}$ of mobile particles on the chain, the spared length per 
particle $\ell_{S}$, and the hopping rate $p$. Measuring all lengths 
in units of $\ell_{0}$, we introduce the dimensionless spared length
\begin{equation}
\bar{\ell}_{S} = \frac{\ell_{S}}{\ell_{0}} \: \: \:.
\end{equation}

It turns out that for all times beyond truly microscopic times $t \alt 2/p$,
the hopping rate $p$ combines with $t$ to yield the time variable
\begin{equation}
\hat{t} = p t \: \: \:,
\end{equation}
which we will use in the sequel. The relation of $\hat{t}$ to 
the Monte Carlo time $t^{(MC)}$ introduces the fit parameter $\tau_{0}$:
\begin{equation}
\hat{t} = \tau_{0} t^{(MC)} \: \: \:.
\end{equation}
Also $\rho_{0}$ and $\bar{\ell}_{S}$ combine into a single important 
parameter. The number of particles that passed over a bead on average 
increases with time, and if it is sufficiently large, the discreteness 
of the individual hopping processes becomes irrelevant. The results then 
depend only on the combination $\bar{\ell}_{S}^{2} \rho_{0}$. 
In our previous work on the motion of individual segments 
\cite{Z15,Z16,Z17}, we found that $\bar{\ell}_{S}$ and $\rho_{0}$ 
separately enter the results only for $\hat{t} \alt 10^{3}$. 
In this time region, a segment on average has moved less than 10 steps 
in the tube and still feels the discreteness of the process. 
For the structure function we find that not even this small time 
region is seriously affected. If we ignore the discreteness of 
the hopping process, the results for all $\hat{t} > 1$ change by less 
than 0.5 \%, which coincides with the accuracy of our numerical evaluation. 
Thus the only microscopic parameters relevant for the coherent structure 
function are the combination $\bar{\ell}_{S}^{2} \rho_{0}$ and the time 
scale $\tau_{0}$. The reason behind this empirically observed suppression 
of initial discreteness effects will be discussed in Subsect.\ 4.C.

We furthermore note that also the discreteness of the underlying chain 
turns out to be unimportant. Provided we normalize the coherent structure 
function by the static structure function calculated for the same 
micro-structure, we for $N \agt 50$ and wave vectors $q \ell_{0} \alt 2$ 
within the above quoted accuracy find the same results for a continuous 
chain as for the model where we sum over discrete bead indices $j,k$. 
This observation is consistent with the independence of statics and 
dynamics discussed in Sect.\ 2.B and illustrates that our calculation 
indeed yields universal results.

Since in our simulations we use the same model as in our previous work 
on segment motion \cite{Z17}, we can take the numerical values of 
the microscopic parameters from there. In analyzing the Monte Carlo data, 
we thus use the value
\begin{equation}
\bar{\ell}_{S}^{2} \rho_{0} = 1.23 \: \: \:.
\end{equation}
Having a much larger set of data available than previously, 
we somewhat readjust the time scale. We use
\begin{equation}
\tau_{0} = 6.8 \cdot 10^{-2} \: \: \:,
\end{equation}
rather than the previous value $\tau_{0} = 6.092 \cdot 10^{-2}$. 
For the logarithmic scale $\log_{10} \hat{t}$ of the figures in 
Ref.~\cite{Z17}, this amounts to a shift by $- 0.048$, which does 
not change the good agreement among theory an experiment which in 
Ref.~\cite{Z17} is shown to hold over about 6 decades of time.

In our formulation of the reptation model, the entanglement length 
$N_{e}$, or the tube diameter $\ell_{0} N_{e}^{1/2}$, equi\-valently, 
do not show up explicitly. They are hidden in the spared length and 
the density of the particles, i.e., in the overall mobility of the chain. 
In contrast, the previous approaches explicitly involve these parameters. 
As discussed at the end of the last subsection, our theory in appropriate 
limits reproduces the previous results, which allows us \cite{Z3} 
to relate our parameters $\bar{\ell}_{S},\rho_{0},p$ to those of these 
other models. Specifically, we find a relation for the entanglement length:
\begin{equation}
N_{e} = 3 \bar{\ell}_{S}^{2} \rho_{0} \: \: \:.
\end{equation}
With the numerical value (4.6) this yields $N_{e} = 3.69$ and implies 
that for the Evans-Edwards model with the smallest possible obstacle 
spacing it needs about 4 steps before the obstacles come into play seriously.

We note that $\bar{\ell}_{S} \rho_{0}$ and thus $N_{e}$ have been determined \cite{Z17} by fitting data for the motion of individual segments in the time range $\hat{t} > 10^{3}$. $N_{e}$ is thus not influenced by the deviations from the full reptation model occuring in the initial range $\hat{t} \alt 10^{3}$, as discussed below, (Sect.\ 5). Thus consistency of the analysis clearly enforces use of the same value $N_{e}$ for all observables. We estimate the uncertainty of this value to be in the range of 5 \%.

We now also can give a quantitative definition of the time scales. 
Identifying the equilibration time $T_{2}$ with the Rouse time of a 
free chain of $N$ segments we from the asymptotic relations among 
the models find
\begin{equation}
\hat{T}_{2} = p T_{2} = \frac{(N+1)^{2}}{\pi^{2}} \: \: \:.
\end{equation}
The reptation time can be defined as the average life time of the original tube
\begin{equation}
\hat{T}_{3} = p \int_{0}^{\infty}\:d t_{0} t_{0} {\cal P}^{*} (t_{0})~.
\end{equation}
Here $d t_{0} {\cal P}^{*} (t_{0})$ is the probability that the tube 
is finally destroyed within time interval $[t_{0},t_{0} + d t_{0}]$. 
It is related to the distribution ${\cal P}^{*}(j_{0},t_{0}|m)$ 
introduced in Eq.\ (4.2) via
\begin{equation}
{\cal P}^{*} (t_{0}) = 
\sum^{N}_{j_{0} = 0} \left[ {\cal P}^{*}(j_{0},t_{0}|0) 
+ {\cal P}^{*}(j_{0},t_{0}|N)\right] \: \: \:.
\end{equation}
Evaluation in the asymptotic limit $N \rightarrow \infty$ yields
\begin{equation}
\hat{T}_{3}^{(\infty)} = \frac{N^{3}}{4 \bar{\ell}_{S}^{2} \rho_{0}} 
= \frac{3}{4} \frac{N^{3}}{N_{e}} \: \: \:.
\end{equation}
In this limit, $\hat{T}_{3}$ is related to the reptation time 
$\tau_{d}$ introduced for the primitive chain model \cite{Z2} as 
\begin{equation}
\hat{T}_{3}^{(\infty)} = \frac{\pi^{2}}{4}\:p \tau_{d} \: \: \:,
\end{equation}
which is the relation used in Sect.\ 3 in our analysis 
of the primitive chain model. 

\subsection{Qualitative discussion of the coherent structure function}

Tube destruction, tube length fluctuations, and internal relaxation 
are different facets of the same microscopic dynamics. 
Therefore any separation of these processes is to some extent artificial. 
Yet some rough discussion based on these concepts is useful with regard 
to the interpretation of the data presented in Sect.\ 5.

We first consider tube length fluctuations $\Delta L$. Here a precise 
definition is possible by identifying $\Delta L$ with the fluctuations 
of the total spared length. On average the spared length equals 
$\ell_{S} \rho_{0} N$, where $\rho_{0} N$ is the average number of 
particles on the chain. Since the fluctuations of the particle number 
essentially are Gaussian distributed we find
\begin{equation}
\frac{\Delta L}{\ell_{0}} \approx \bar{\ell}_{S} \sqrt{\rho_{0} N} 
= \sqrt{\frac{1}{3} N_{e} N} \: \: \:,
\end{equation}
where $N_{e}$ is introduced via Eq.\ (4.8). Within the equilibration 
time $T_{2}$ tube length fluctuations on average on both chain ends 
replace a piece of length $\Delta L/2$ of the original tube by a new 
piece. In real three dimensional space, this new piece has an average 
extension of $\sqrt{\ell_{0} \Delta L/2}$, and to observe the effect, 
the scattering vector must obey $q^{2} \ell_{0} \Delta L/2 \agt 1$, or
\begin{equation}
q^{2} \ell_{0}^{2} \agt \sqrt{\frac{12}{N_{e} N}} \: \: \:,
\end{equation}
equi\-valently. For large wave vectors, the coherent structure function 
$S_{c}(q,t;N)$ is determined by the part of the original tube that 
still exists at time $t$. Taking only tube length fluctuations into 
account, we therefore find for $t \approx T_{2}$:
\begin{eqnarray}
\bar{S}_{c} (q,T_{2};N) &=& \frac{S_{c}(q,t;N)}{S_{c}(q,0;N)} 
\approx \frac{N - \Delta L/\ell_{0}}{N}
\nonumber \\
&\approx& 1 - \sqrt{\frac{N_{e}}{3 N}} \: \: \:,
\end{eqnarray}
provided $q$ is large compared to the bound (4.15). This estimate 
yields the effect of fully developed tube length fluctuations. 
With increasing chain lengths it clearly is suppressed, but it should 
be well visible up to fairly long chains $N/N_{e} \approx 10^{2}$.

Tube destruction in the sense of the primitive chain model is 
responsible for the main part of the decay of the coherent structure 
function. According to Eq.\ (4.16), it is the dominant effect as soon 
as $1 - \bar{S}_{c}(q,t,N) \gg \sqrt{N_{e}/3 N}$. 
For $t \rightarrow T_{2}$, it is strongly correlated with tube 
length fluctuations which, seen from a microscopic point of view, 
drive the tube destruction. It therefore is no surprise that evaluating 
the coherent structure function of the primitive chain model, 
$\bar{S}_{DE} (q^{2} R_{g}^{2},t/T_{3})$, for $q^{2} R_{g}^{2} \gg 1$ 
and time $t = T_{2}$, we find a decrease of the same order of magnitude 
as that due to tube length fluctuations.
\begin{eqnarray}
\lefteqn{
1 - \bar{S}_{DE} \left(q^{2} R_{g}^{2},\frac{T_{2}}{T_{3}}\right) 
\stackrel{ q^{2} R_{g}^{2} \gg 1} {=}  1 - \bar{S}^{(c)} (T_{2},N)}
\nonumber \\ 
& & = \sqrt{\frac{2}{\pi} \frac{T_{2}}{T_{3}}} + O 
(e^{-\mbox{\footnotesize const}\:T_{3}/T_{2}}) \approx 
\sqrt{\frac{8}{3\pi^{3}} \frac{N_{e}}{N}} 
+ O (e^{- \mbox{\footnotesize const}\:N}) \:.
\nonumber\\
\end{eqnarray}
Recall that $\bar{S}^{(c)}(t,N)$ (Eq.\ (3.6)) is the limit of 
$\bar{S}_{DE}$ for $q^{2} R_{g}^{2} \gg 1$. Comparison of Eqs.\ (4.16), 
(4.17) illustrates that tube length fluctuations cannot properly be 
separated from tube destruction. Being of the same order of magnitude 
for $t \approx T_{2}$, they strongly interfere in the full theory.  

We finally note that the use of $T_{3}$ as fit parameter, necessary 
to fit the results of the primitive chain model to our data, illustrates 
that tube length fluctuations influence the efficiency of tube destruction 
for all times \cite{Z6}, even for chains of lengths $N/N_{e} \approx 100$.

Internal relaxation can be analyzed by considering a subchain of length 
$M$ in the center of an infinitely long chain. We thus consider 
the coherent structure function
\begin{equation}
S_{c}(q,t;M,\infty) = \sum^{M}_{j,k=1} \left\langle 
\overline{e^{i {\bf q}({\bf r}_{j}(t) - {\bf r}_{k}(0))}}
\right\rangle^{\infty} \: \: \:.
\end{equation}
Within time $t$, segment $j$ is displaced along the tube by 
$\bar{\ell}_{s} n$ steps, and since the tube has a random walk 
conformation, the average over the paths of segment $j$ yields 
\begin{eqnarray*}
\left\langle e^{i {\bf q}({\bf r}_{j}(t) - {\bf r}_{k}(0))} 
\right\rangle^{\infty} = e^{- \bar{q}^{2} |j + \bar{\ell}_{s} n - k|}~.
\end{eqnarray*}
This expression is to be averaged over the distribution 
${\cal P}_{1}(n,t)$ of the number $n$ of particles that diffused 
over bead $j$. We note that in the situation considered here 
${\cal P}_{1}(n,t)$ is independent of $j$, since the process 
is translationally invariant along an infinitely long chain. We thus find
\begin{equation}
S_{c}(q,t;M,\infty) = \sum^{+ \infty}_{n = - \infty}\:{\cal P}_{1} (n,t) 
\sum^{M}_{j,k=1} e^{- \bar{q}^{2} |j + \bar{\ell}_{s} n - k|}
\: \: \:.
\end{equation}
Now shifting $j + \bar{\ell}_{S} n \rightarrow j$, 
we can rewrite this expression as 
\begin{eqnarray}
S_{c}(q,t;M,\infty) &=& \sum^{+ \infty}_{n = - \infty}\:{\cal P}_{1} (n,t) 
\Bigg\{ \sum^{M}_{j,k=1} e^{- \bar{q}^{2} |j - k|}
\nonumber \\
&+& \sum^{M}_{k=1} \left[ \sum^{M+\bar{\ell}_{s} n}_{j = M+1}\:
e^{- \bar{q}^{2} |j-k|} - \sum^{\bar{\ell}_{s} n}_{j=1}\: 
e^{- \bar{q}^{2} |j-k|}\right] \Bigg\}
\nonumber\\
\end{eqnarray}
Since ${\cal P}_{1} (n,t)$ is normalized to $1$, the first term 
in Eq.\ (4.20) yields the static structure function of the subchain. 
All time dependence results from the motion of the end pieces of the 
subchain and thus is due to the length fluctuations and the shift of 
the subchain in the tube. Proper interior relaxation within the subchain 
has no effects on the coherent structure function.

We want to stress that this does {\em not} imply that our theory 
{\em neglects} interior relaxation. The basic relation (4.19) only 
exploits the fact that the segment $j$ has carried out a random walk 
of $\bar{\ell}_{S} n$ steps. It does not imply that this walk is along 
the precise original configuration of the chain. Rather the path will 
have changed by the motion of other segments, i.e., internal relaxation. 
What our result shows is that the effects of interior relaxation 
{\em average out} in the structure function as calculated within 
the reptation model. This also explains the observation in 
Subsect.\ 4.B that the discreteness of the individual steps 
is unimportant numerically. It yields only a correction to the 
second term in Eq.\ (4.20), which by itself is very small in the 
time range $\hat{t} \alt 10^{3}$ where discreteness corrections 
might show up. Deviations among theory and data which we will find 
for short times and wave vectors of the order of the inverse segment 
size therefore must originate from effects not taken into account 
in the reptation model.

In this context some further remark on the model of Rouse motion 
in a coiled tube seems appropriate. Concerning the {\em dynamics} 
along the tube, our hopping model is essentially equi\-valent to Rouse 
motion. Specifically, beyond microscopic times the motion of individual 
segments approaches Rouse type motion, as has been shown in Ref.~\cite{Z16}, 
Sect.\ 3. The difference found for the structure function results from 
the imposed initial conditions, only. It reflects the fact that 
the structure function measures 
correlations between two segments $j,\;k$ and thus is influenced 
by the initial configuration of the subchain connecting $j$ with $k$.    

To summarize, we have seen that in the limit $N \rightarrow \infty$ 
with $q^{2} R_{g}^{2} \sim q^{2} N$ kept fixed, the coherent structure 
function of the total chain varies only on scale $T_{3}$. In this limit, 
$\bar{S}_{c}(q,t;N)$ is given by the primitive chain model. 
For shorter chains, tube length fluctuations come into play. 
They change the effective reptation time and lead to an initial 
decrease of $\bar{S}_{c}$ on time scale $T_{2}$. Both effects are 
important for chains up to lengths $N/N_{e} \approx 10^{2}$, at least. 
Proper internal relaxation, however, averages out. 

With regard to the data presented in Sect.\ 5, we want to close 
this subsection with a brief discussion of the scattering 
$S_{c}(q,t;M,N)$ from the central subchain of length $M \ll N$ 
in a chain of finite length $N$. For small times (and sufficiently 
large wave vectors), we expect to see small fluctuations of the length 
of the part of the tube occupied by the subchain. Since, however, these 
fluctuations do not lead out of the original tube, they are much less 
efficient in reducing the scattering function than the tube length 
fluctuations relevant for the total chain: on average, up to times 
of the order of the relaxation time $T_{2}(M)$ of the subchain, 
its position in the tube has not changed seriously. A definite 
decrease of $S_{c}(q,t;M,N)$ sets in for times large enough for 
the subchain to leave its original part of the tube. For very large 
chain lengths $N$, this may occur for times $t \ll T_{2} (N)$. 
$S_{c}$ then decreases like $t^{-1/4}$, this law reflecting the 
$t^{1/4}$-law (Eq.\ (1.1)) for the motion of an internal segment. 
For $T_{2}(N) \ll t \ll T_{3}(N)$, the motion of the subchain is 
driven by particles created by the chain ends which leads to a decrease 
like $t^{-1/2}$ (cf.\ Eq.\ (1.1)). Finally, of course, this behavior is 
cut off by tube destruction reaching the position of the subchain. 
On the quantitative level this behavior can be extracted from our theory. 
In the discussion of our data we in some figures will include curves 
which show the effect of relaxation of the (sub-) chain moving in 
an infinitely long tube, i.e., neglecting all tube renewal effects. 
                
\subsection{Approximate analytical parameterization of our results 
for the total chain}

The integral equations which yield the normalized coherent structure 
function $\bar{S}_{c}(q,t;N)$ of the full reptation model, can be solved 
only numerically. Furthermore, also the kernel and the inhomogeneity 
in these equations do not take a simple form \cite{Z3}. In particular, 
the inhomogeneity is given by some integral with a lengthy integrand. 
Thus, to present the results in a more easily tractable form some more 
or less heuristic analytical parameterization might be useful.

It should be clear that such a parameterization cannot be very simple. 
The coherent scattering function involves two time scales: $T_{2}$ and 
$T_{3}$. Furthermore also wave vector and chain length combine into 
two relevant variables: $q^{2} R_{g}^{2}$ and 
$q^{2} \ell_{0}^{2} \sqrt{N_{e} N} \sim q^{2} R_{g}^{2} \sqrt{N_{e}/N}$. 
The latter variable should govern the contribution of tube length 
fluctuations (cf.\ Eq.\ (4.15)). In searching for a reasonable 
parameterization we exploit the fact that for very long chains 
our results reduce to those of the primitive chain model. Furthermore, 
as shown in Sect.\ 3.A, the functional form of $\bar{S}_{c}$, given by 
this model, can be used to fit the long-time tail also for non-asymptotic 
chain lengths. This suggests to use an ansatz similar to de Gennes' form (3.9):
\begin{equation}
\bar{S}_{\scriptsize \mbox{Fit}} (q,t;N) 
= (1 - B) H (q,t;N) + B \bar{S}_{DE} 
\left(q^{2} R_{g}^{2},\frac{\hat{t}}{\hat{T}_{3, \scriptsize \mbox{Fit}}}
\right)
\end{equation}
The function $\bar{S}_{DE} (Q,\tau)$ is given by Eqs.\ (3.2)-(3.5). 
Analyzing the numerical results of our theory in a large range of 
wave vectors and chain lengths which exceeds that used in the simulations, 
we find that the effective reptation time weakly depends on $q$. 
We use the expression
\begin{eqnarray}
\lefteqn{\hat{T}_{3, \scriptsize \mbox{Fit}} = \frac{3}{4} N_{e}^{2}
\left(\frac{N}{N_{e}}\right)^{3}\;\cdot}
\\ 
&&\left\{ 1 - \left[ \frac{2.1}{(9 + N/N_{e})^{0.5}} 
- - \frac{2.5}{9 + N/N_{e}}\right] \cdot 
\left( 1 + \frac{0.25}{(1 + 0.3 q^{2} R_{g}^{2})^{2}}\right)\right\}.
\nonumber
\end{eqnarray}
The leading term gives the asymptotic law (4.12), the first correction 
is of order $\sqrt{N_{e}/N}$, in agreement with a result by Doi \cite{Z6}. The prefactor of this correction differs somewhat from Doi's result. However it must be noted that eqs. (4.21) - (4.25) are meant to reproduce the scattering function in a wide parameter range. Thus eq. (4.22) should not literally be compared to Doi's result. 

We now turn to the coefficient $B$, which is taken as  

\begin{equation}
B = \exp \left\{ - \left( \frac{q^{2} R_{g}^{2} 
\sqrt{\frac{N_{e}}{N}}}{1 + q^{2} R_{g}^{2} 
\sqrt{\frac{N_{e}}{N}}}\right)^{0.6} \frac{2.8}{2.4 
+ \sqrt{N/N_{e}}}\right\} \: \: \:.
\end{equation}
This form is consistent with the discussion of the influence of 
tube length fluctuations given in the previous subsection. 
$B$ tends to 1 for $q^{2} R_{g}^{2} \sqrt{N_{e}/N} \ll 1$ and 
yields $B \approx 1 - O (\sqrt{N_{e}/N})$ in the limit 
$N \rightarrow \infty$, $q$ fixed.

The initial decrease of the scattering function over a sizeable range 
of time is approximated by a stretched exponential with an effective 
exponent which strongly decreases with increasing $q^{2} R_{g}^{2}$. We take 
\begin{equation}
H (q,t,N) = \exp \left\{ - \left( \frac{\hat{T}_{2}}{\hat{T}_{3, 
\scriptsize \mbox{Fit}}}\right)^{x_{1}} \left[ 3.2 \left( 
\frac{q^{2} R_{G}^{2}}{0.7 + q^{2} R_{g}^{2}}\right)^{2} 
\frac{\hat{t}}{\hat{T}_{2}}\right]^{x_{2}} \right\}
\end{equation}
\begin{eqnarray}
x_{1} &=& 0.1 + \frac{0.9}{1 + 0.2 q^{2} R_{g}^{2}}
\nonumber \\
x_{2} &=& 0.25 + \frac{0.75}{1 + 0.1 q^{2} R_{g}^{2}} \: \: \:.
\end{eqnarray}  
We recall Eq.\ (4.9): $\hat{T}_{2} = (N + 1)^{2}/\pi^{2}$. 

The parameterization (4.21)-(4.25) is fitted to the numerical evaluation 
of our theory in the range $0.1 \leq q^{2} R_{g}^{2} \alt 50$; 
$10 \alt N/N_{e} \alt 340$; $0 \leq \hat{t} < \infty$. In most of 
that range it reproduces our full results within deviations less 
than $0.015$ in absolute value. It becomes worse for shorter chains 
$(N/N_{e} \alt 30)$ and larger wave vectors $(q \ell_{0} \agt 2)$ 
in the large time regime $(\hat{t} \agt 0.1 \hat{T}_{3})$. In this 
region the large-time tail predicted by the full theory deviates 
somewhat from the functional form given by the primitive chain model, 
and also small differences between the continuous and the discrete version 
of the model start to show up. For $N/N_{e} \alt 10$ our theory breaks down, mainly due to the fact that tube length fluctuations become important even for $\hat{t} \approx \hat{T}_{3}$. They thus affect the kernel 
${\cal P}^{*}(j_{0} t_{0}|m)$ of the integral equation (4.2), 
and this effect has not been considered in our theory. 

In the sequel we compare our Monte Carlo data to the results of the 
numerical evaluation of our full theory. It should be noted, however, 
that on the scale of the plots shown the difference between the results 
of the full theory and the effective parameterization (4.21)-(4.25) 
would be almost invisible.            

\section{Comparison of Monte Carlo data to full reptation theory}

The reptation model in full detail treats the motion along the tube axis, 
but it does not account for all relaxation processes in a realistic chain. 
In particular, motions `perpendicular to the tube', like the creation and 
decay of larger side branches, will be present for the Monte Carlo chain 
but are not treated properly in the reptation model. Since larger side 
branches emerge by fusion of hairpins,  they in the framework of our 
model would introduce some kind of interaction of the particles. 
Scattering vectors of magnitude $q \approx 1$ certainly resolve such 
effects which result in some additional relaxation on a microscopic 
time scale. In fitting the data to the theory we therefore allow for 
a phenomenological amplitude $A \leq 1$, multiplying the normalized 
structure function $\bar{S}_{c}(q,t;N)$ calculated for pure reptation. 
We typically find values $A \approx 0.9$ for $q = 1$, $A \approx 0.98$ 
for $q = 0.5$, and $A \approx 1$ for all smaller $q$ measured. 
Thus the deviations among theory and data which are found for small times, 
rapidly decrease with decreasing $q$.   

For a first illustration, we in Fig.~7 show our results for 
$N = 80,\; q = 0.5024\; (q^{2} R_{g}^{2} = 3.32)$. 
Data points result from runs extending to $10^{5}$ or $10^{8}$ 
Monte Carlo steps. The full curve gives the result of our theory, 
with the amplitude adjusted to $A = 0.985$. Long dashes represent a 
fit to the primitive chain model, with $B_{DE}$ and $\hat{T}_{3} = 
\tau_{0} T_{3}^{(MC)}$ taken from Table 1. The short dashed curve 
results by neglecting tube destruction and is calculated as the 
structure function of a chain of length $N$ in an infinitely long tube. 
Since the primitive chain model takes care only of tube destruction, 
the difference among the two dashed curves is due to tube length 
fluctuations. Thus Fig.~7 demonstrates that the tube length 
fluctuations strongly influence the structure function up to 
times $t \approx T_{2}$. For much larger times tube destruction 
is the dominant process. 

These results are typical for all chain lengths and wave vectors. 
Figs.\ 8 and 9 show our results for $N = 160$ and $320$, respectively. 
Beyond some short time regime, we find excellent agreement between theory 
and data, with only the amplitude $A = A (q,N)$ taken as adjustable 
parameter.  Comparing Fig.~9 to Fig.~4, (note the difference of times 
scales: $\log_{10} \hat{t} = \log_{10} t^{(MC)} - 1.167)$, we again see 
the effect of tube length fluctuations: for $N = 320$, the primitive chain 
model can fit the data only for $t \agt 20 T_{2}$. For smaller times, 
tube length fluctuations become relevant, their effect increasing with 
increasing $q$, as expected. 
For $N = 160, q = 0.1\: (q^{2} R_{g}^{2} \approx 0.267)$, scattering 
cannot resolve the structure of the tube and the effect of tube length 
fluctuations is suppressed. The corresponding curve shown in Fig.~8 
essentially coincides with the result of the primitive chain model, 
provided we adjust the effective time scale $T_{3}$. 

To gain additional information on the chain dynamics, we also measured 
and calculated the structure function of the central piece of length 
$M$ in a chain of length $N$. Fig.~10 shows our results for 
$M = 80,\; N = 320,\; q = 0.5$. With these parameter values 
it is directly comparable to Fig.~7, where $M = N = 80,\; q = 0.5024$. 
The central piece evidently is not influenced by tube length fluctuations. 
In contrast to Fig.~7, the decrease of $\bar{S}_{c}(q,t;M,N)$,  
in Fig.~10 must be due to relaxation within the initial tube, 
up to times where tube destruction reaches the subchain measured. 
Indeed, in Fig.~10 the structure function calculated neglecting 
tube destruction (broken line) over a large time range coincides 
with the results of our full theory and simulations. The time 
$T_{R}(M,N)$ it needs for tube destruction to reach the subchain, 
can be estimated from the relation 
\begin{equation}
\bar{\ell}_{S}  n_{\rm max} (T_{R} {\scriptstyle(M,N)}) 
= \frac{1}{2} (N - M)\: \: \:.
\end{equation}
This time is indicated in Fig.~10 and is close to the time where 
the dashed line starts to deviate from our full result. Recall from 
Sect.\ 4.A  that $\ell_{S} \overline{n_{\rm max}(t)}$ gives 
the average length of the end-piece of the tube that has been 
destroyed up to time $t$.

As pointed out in Sect.\ 4.C, the decrease of $\bar{S}_{c}(q,t;M,N)$ 
seen in Fig.~10 in the range $t < T_{R} (M,N)$ is not due to relaxation 
within the subchain $M$ considered, but results from the internal 
dynamics of the total chain which leads to fluctuations of the position 
of the subchain in the original tube. This is illustrated in Fig.~11, 
which for given $N = 320$ and $q = 1.0$ compares the scattering from 
central subchains of lengths $M = 40,\;80,\;160,\;320$. 
We note some peculiar behavior: as function of time, 
the normalized scattering function of the total chain 
$M = N = 320$ initially decreases faster and eventually crosses results for internal pieces. 
This shows the efficiency of tube length fluctuations in comparison 
to the internal dynamics of the chain. According to the discussion of Sect.\ 4.C, subchains 
$M \alt N - \sqrt{\frac{1}{3} N_{e} N} \approx 300$ (for $N = 320)$ 
are not influenced by tube length fluctuations, and for a large time 
range scattering from such subchains is determined by the diffusive 
motion of the total subchain in the initial tube. Since the diffusion 
coefficient of this motion scales as $1/M$, the diffusion of the longer 
subchains is very slow. Taking place in the initial tube, it is also less 
efficient in reducing $\bar{S}_{c}$ than tube length fluctuations or 
the tube destructing shift of the total chain. For these reasons, 
the normalized coherent scattering for $M = 160$ stays above the result 
for the total chain up to times $t \approx T_{3}\left(N\right)$. For larger times the theory predicts that as for the shorter subchains the normalized scattering from subchain $M = 160$ falls below the scattering from the total chain, but the effect is too small to be visible in Fig.~11, and it clearly is not resolved by the accuracy of the data. We also note that tube destruction reaches the initial position of the subchain $M = 160$ only for $p T_{R} (160,320)\approx 10^{5.8}$. 
With decreasing $M$, diffusion in the tube becomes faster, and for the 
shortest subchain, $M = 40$, we see indications of a new regime: 
the slope of the curve seems to change near $\hat{t} = 10^{5.4}$. 
A doubly logarithmic plot reveals that in the range 
$10^{5.4} \leq \hat{t} \leq 10^{6.1}$ the data approach a 
$\hat{t}^{-1/2}$-law, and the upper bound of this interval is close 
to the time $p T_{R} (40,320) \approx 10^{6.4}$. These observations 
are consistent with the discussion of Sect.\ 4.C. 

Finally we consider the deviations between theory and data present 
for $\hat{t} \alt 10^{3}$ and large wave vectors. A first important 
observation is that these deviations roughly are of the same size  
for interior pieces of the chain as for the total chain, see Fig.~11. 
Therefore their dominant source cannot be searched in our treatment of 
tube length fluctuations in the mean hopping rate approximation. 
To show this in more detail, we in Fig.~12 have plotted the deviation 
between data and theory for wave vector $q = 1.0$ and all chains and 
subchains measured. We note some end-effect: for the total chain 
$N = 160 = M$ the deviation is about 20 \% larger than for the 
central piece $M = 160$ in the chain $N = 320$. However, the bulk 
of the effect is independent of $N > M$, as illustrated by the curves 
for $M = 40$ and $M = 80$, which each in fact combine data for 
$N = 160$ and $N = 320$. For these internal pieces, tube length 
fluctuations are irrelevant, and in the time range shown, our evaluation 
of the reptation model is exact within the accuracy (0.5 \%) of our 
numerics. Consequently the initial deviations must be dominated by some 
relaxation in the interior of the subchain $M$ which is not taken into 
account by the reptation model. We note that the amplitude $A$ of 
the effect depends on $M$ only weakly, but strongly increases with $q$, 
as is illustrated by Figs.\ 8 and 9. It roughly is proportional to $q^{2}$. 
These observations are consistent with an interpretation as 
micro-structure effects, which clearly increase strongly with the resolving power of the experiment and are expected to show a weak dependence $\sim 1/M$ on (sub-)chain length.

In our previous work on the motion of individual segments, we found 
initial transients which also decayed on a time scale 
$\hat{t} \alt 10^{3}$. These could be traced back to the discreteness 
of both the hopping process and the underlying chain and could be 
largely explained by a fully discrete analysis of our model. 
As mentioned in Sect.\ 4.A, it turns out that for $\bar{S}_{c}(q,t,M,N)$ 
this discreteness numerically is irrelevant. In the light of 
the discussion of Sect.\ 4.C, this is quite understandable: 
the internal dynamics of the subchain averages out. We thus have 
to search for effects going beyond our implementation of the reptation 
model. One source of discrepancy immediately comes into mind. 
A particle sitting on a bead is no exact representation of a hairpin. 
It, for instance, is not clear whether the bead sits in the tip or in 
the base of the hairpin. Simple estimates of such effects easily yield 
a correction amplitude $1-A$ of the desired order of magnitude. However, 
the dynamics of single hairpins cannot explain a time scale of order 
$10^{3}$ since it essentially relaxes within one Monte Carlo step. 
Slower relaxation occurs for larger side branches. Already a structure 
created by fusion of two hairpins shows a relaxation time of the order 
of $10 MC$ steps. Indeed, the fusion of hairpins destroys mobile units 
of the Monte Carlo chain. In the language of our model this induces some 
dynamical interaction of the particles. We therefore believe that the 
observed initial transients are due to the interference of reptation 
with the dynamics of larger side branches.

\section{Conclusions}

In this paper, we presented results of simulations of the Evans-Edwards model, measuring the coherent structure function of a reptating chain in a large range of time and wave vector. This model contains the full reptational dynamics of a single chain, including internal relaxation, tube length fluctuations, and tube renewal, but omits all the dynamics of the environment. In particular, it ignores constraint release. We compared the data to two simplified versions of the reptation model as well as to our recent evaluation of the full theory. We found that the primitive chain model proposed by Doi and Edwards \cite{Z4,Z2} yields a good fit to the data for times large compared to the Rouse time $T_{2}$, provided we allow for a phenomenological amplitude factor and treat the reptation time $T_{3}$ as a second adjustable parameter. De Gennes' approach \cite{Z5} which adds a contribution meant to take one dimensional internal relaxation along the tube into account, cannot explain the data consistently. So the success of this approach in fitting neutron scattering data on polymer melts \cite{Z21,Z22,Z23} indicates that the local contribution $S^{(\ell)}(q,t)$ to De Gennes' structure function phenomenologically accounts for some dynamics not contained in the reptation model. In view of our findings for the Evans-Edwards model we would search the origin of the observed initial decay in a coupling of reptation to motion perpendicular to the tube. Since in the experimental time range the initial decay accounts for most of the variation of the experimental curves \cite{Z23}, we have not tried to fit these curves with our theory.

Provided we allow for a single adjustable amplitude factor we, beyond the short time range influenced by microstructure relaxation, find quantitative agreement between our simulation data and our analytical evaluation \cite{Z3} of full reptation theory. The theory in all details explains not only the scattering function of the total chain but also the scattering from interior subchains. Except for the phenomenological amplitude, which notably differs from $1$ only for wave vectors of the order of the inverse segment size, all parameters of the theory essentially are fixed by our previous analysis \cite{Z16,Z17} of the motion of internal segments. Thus our version of the reptation model consistently explains both details of the internal motion and global features like the scattering function.

Our analysis shows that the effects of internal relaxation contained in the reptation model and seen, for instance, in the motion of internal segments, for the scattering functions average out. In the reptation model, the time dependence of the scattering functions is determined solely by the interplay of tube destruction and tube length fluctuations. The latter not only determine the scattering function up to times of order $T_{2}$, but also for intermediate chain lengths lead to the decrease \cite{Z6} of the reptation time $T_{3}$ in comparison to the asymptotic law, clearly seen also in the scattering function. Numerically evaluating our theory, we found these effects to be well visible up to chain lengths of the order of 300 entanglement lengths. This is obvious from the numerical parameterization of our results, given in Section 4~D.

Our theory deviates from the results of the simulation in the initial time region $\hat{t} \alt 10^3$. We stress that this is not due to the approximations inherent in our treatment of tube destruction. Deviations of the same order of magnitude occur for the total chain and for internal subchains, and the coherent structure function of internal subchains in the initial time region is not affected by our approximations. Thus the need to introduce some phenomenological parameter to take short time relaxation into account illustrates that even for the Evans-Edwards model, which is the most accurate computer-experimental implementation of the reptation model we can think of, some relaxation 'perpendicular' to the tube shows up. Clearly, for a melt, such effects must be much larger. However, our analysis suggests some way to suppress these contributions. The scattering from a chain of length $N=M$ should be compared to the scattering from the central piece of length $M$ in a chain of length $N \approx 2 M$. Up to small end effects, the difference of the two coherent scattering functions should be due to reptation alone, and can be calculated from our theory. To observe an effect, the experiments must cover a time range extending at least up to the Rouse time, with chain length being of the order $N/N_{e} \agt 20$. Unfortunately, for physical experiments, this seems to be out of reach. As shown by the work of Ref. \cite{Z10} it might be feasible in simulations.
\\ \\
{\bf Acknowledgment:}
This work was supported by the Deutsche Forschungsgemeinschaft, SFB 237.

%\newpage

\end{multicols}

\newpage 

\begin{table}[H]
\begin{center}
\begin{tabular}{c|c|c|cc} 
$N^{(MC)}$ & $q$ & $\log_{10} T_{3}^{(MC)}(q,N)$ & $B_{DE}  (q,N)$\\ \hline 
80 & 0.5024 & 5.90 & 0.91\\ \hline
160 & 0.10 & 6.91 & 0.997 \\ 
& 0.50 & 6.87 & 0.89\\
& 1.00 & 6.82 & 0.76\\ \hline
320 & 0.25 & 7.85 & 0.965 \\ 
& 0.50 & 7.82 & 0.88 \\
& 1.00 & 7.80 & 0.78 \\ \hline
\end{tabular}
\end{center}
\caption[]{Parameter values for the fit of 
$B_{DE} \bar{S}_{DE} (q^{2} R_{g}^{2},t/T_{3})$ to the data. 
Reptation time $T_{3}^{(MC)}$ is measured in Monte Carlo steps.}
\end{table}

%\newpage

\begin{figure}
\label{fig1}
\begin{center}
\epsfig{figure=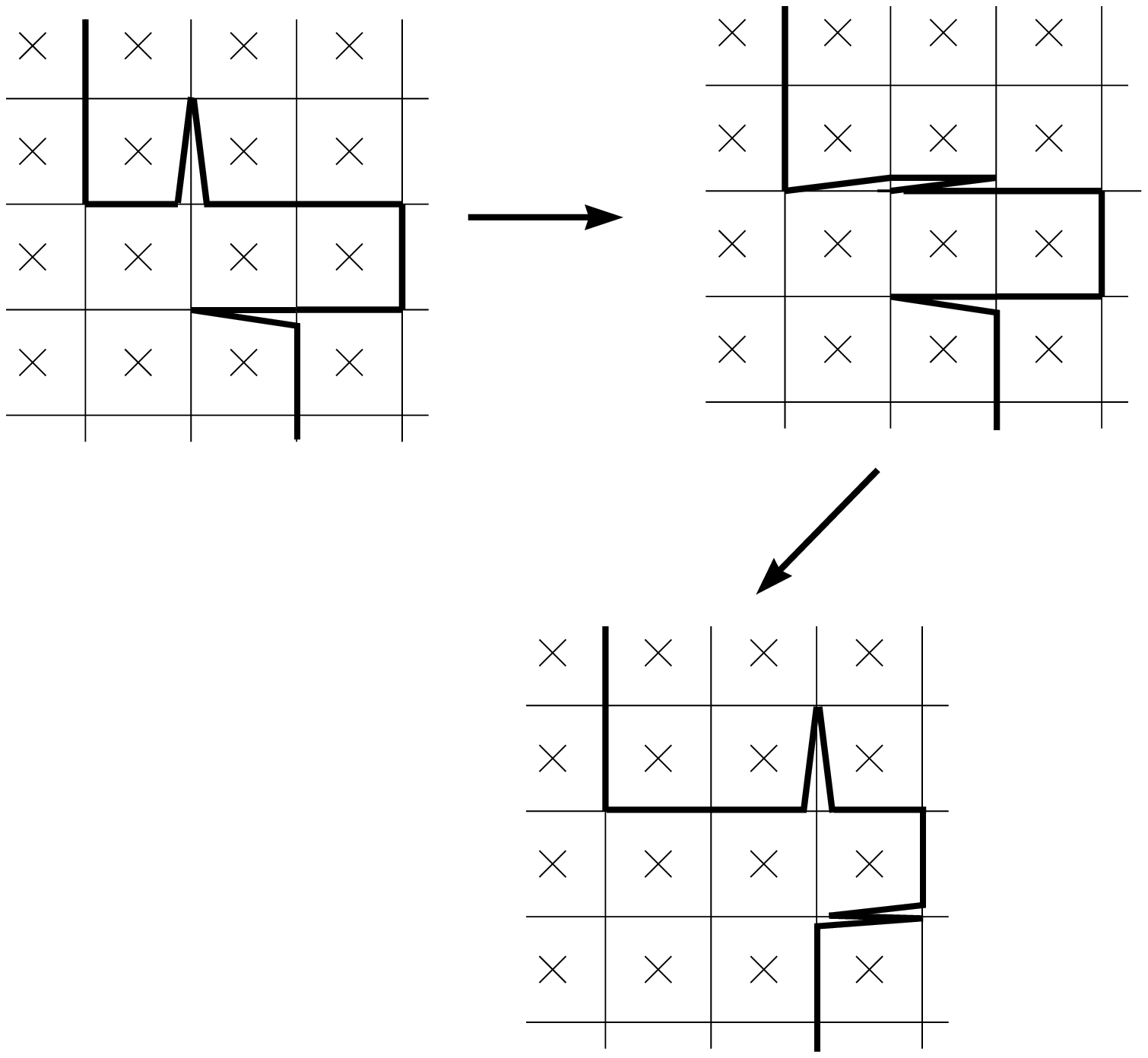,width=0.5\linewidth} 
\caption[]{A series of subsequent chain configurations, illustrating 
the microscopic dynamics of the Evans-Edwards model 
(in its two-dimensional version). 
The crosses represent impenetrable obstacles.}
\end{center}
\end{figure}

\begin{figure}
\label{fig2}
\begin{center}
\epsfig{figure=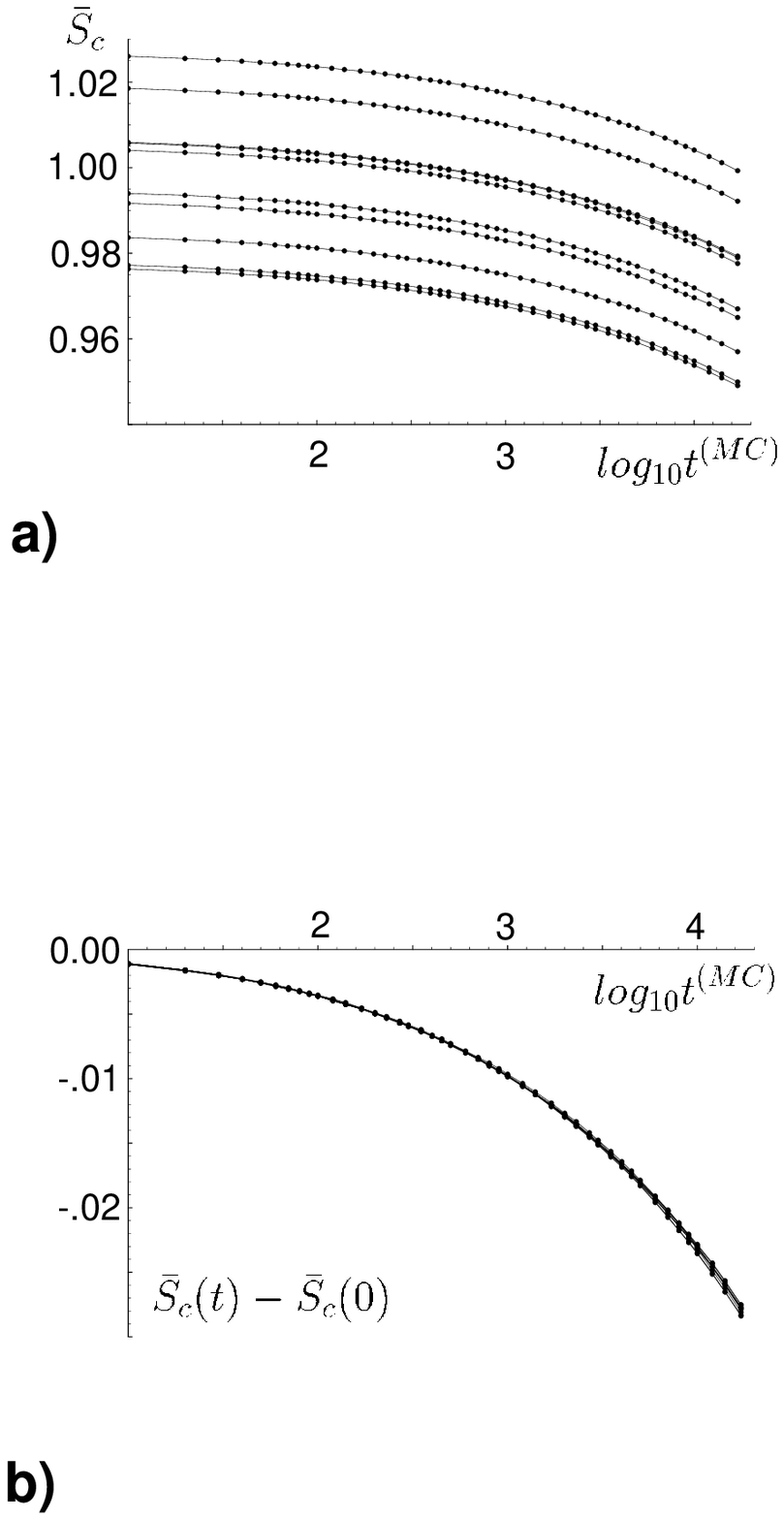,width=0.5\linewidth} 
\caption[]{Data for the coherent structure function $(N = 320,~q = 0.5)$ 
as function of $\log_{10} t^{(MC)}$ in the short time regime. 
$\bar{S}_{c} (q,t^{(MC)};N)$ is normalized to the exact static 
structure function $S_{0}(q,N)$ (Eq.\ (2.6)). ~
a) 10 sets of data, each averaged over $10^{3}$ independent runs. 
The lines serve to guide the eye. ~
b) The same sets of data as in a), but with $\bar{S}_{c}(q,0;N)$ subtracted.}
\end{center}
\end{figure}

\begin{figure}
\label{fig3}
\begin{center}
\epsfig{figure=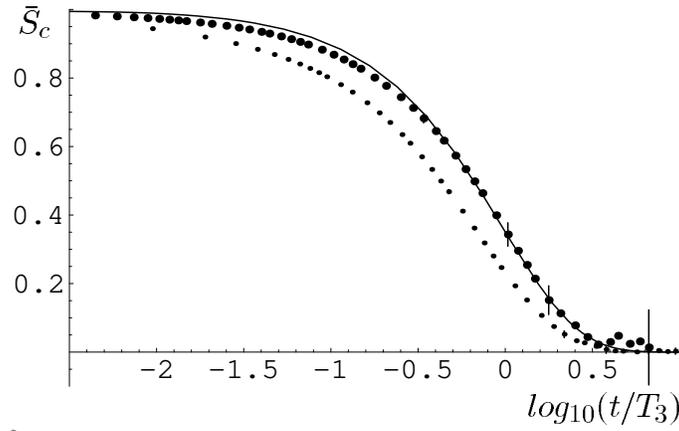,width=0.5\linewidth} 
\caption[]{$\bar{S}_{c}$ measured for $q^{2} R_{g}^{2} \approx 3.3$ 
as function of $t/T_{3}$ in the long time regime. 
Heavy dots: $N = 320$; small dots: $N = 80$. 
Full curve: primitive chain result (3.3). $T_{3} (N = 320)$ has been 
adjusted to bring the long-time tail of the data for $N = 320$ on top 
of the theoretical curve. $T_{3} (80) = T_{3} (320)/4^{3}$. 
Some typical error bars (two standard deviations) are shown 
for $N = 320$. The statistical error rapidly decreases with 
decreasing $t/T_{3}$. For $t/T_{3} < 10^{-0.5}$ it is smaller 
than the size of the dots.}
\end{center}
\end{figure}
\newpage

\begin{figure}
\label{fig4}
\begin{center}
\epsfig{figure=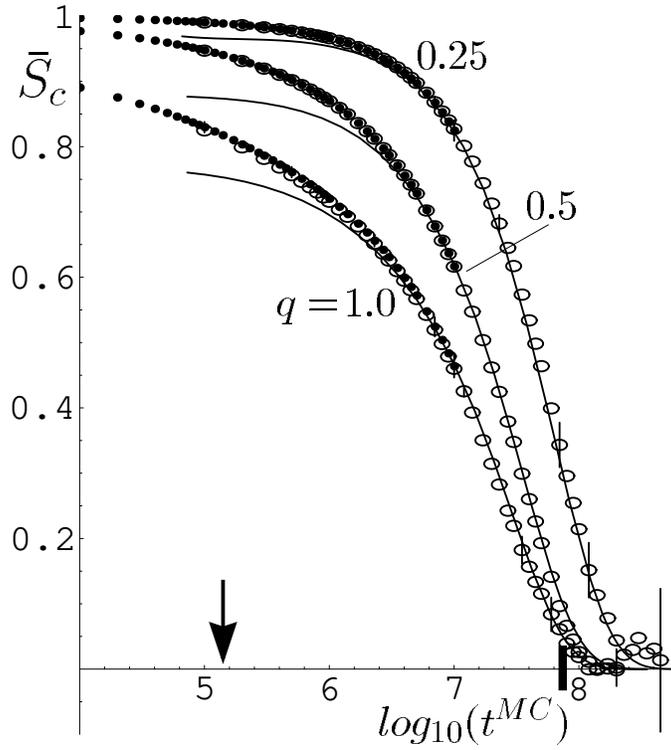,width=0.5\linewidth} 
\caption[]{Data for $\bar{S}_{c}(q,t,N)$ for chain length $N = 320$, 
wave vectors $|{\bf q}| = 0.25,~ 0.5, ~1.0$, as function of the Monte Carlo 
time; dots: medium time runs $(t_{\rm max}^{(MC)} = 10^{8})$; 
circles: long time runs $(t_{\rm max}^{(MC)} = 4.5 \cdot 10^{9})$. 
The curves give the results of the primitive chain model, fitted to 
the data as explained in the  text. The arrow points to the equilibration 
time $T_{2}^{(MC)}(320)$, the heavy bar gives 
$T_{3}^{(MC)}(320) = 10^{7.82}$. Some typical error bars 
(two standard deviations) are shown for $q  = 1.0, ~0.25$.}
\end{center}
\end{figure}
\newpage

\begin{figure}
\label{fig5}
\begin{center}
\epsfig{figure=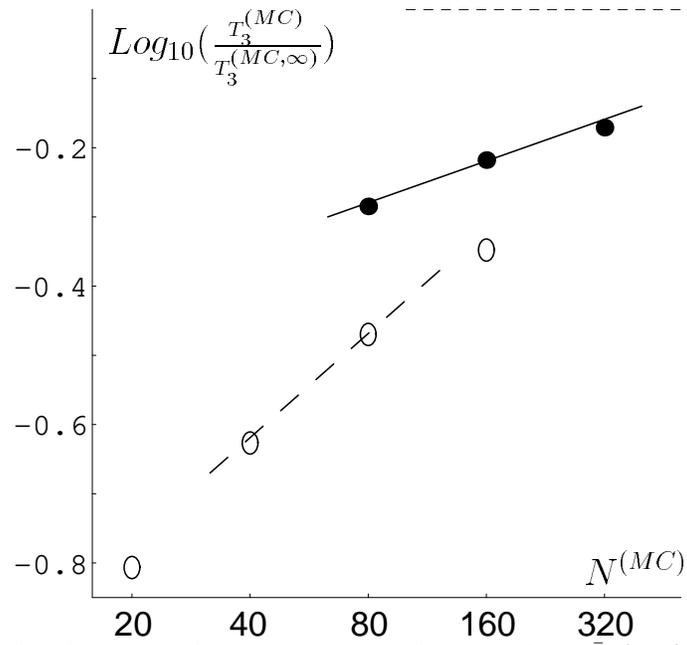,width=0.5\linewidth} 
\caption[]{Results for the reptation time defined in terms of the long time behavior of $\bar{S}_{c}$ (dots) or in terms of the motion of the endsegment (circles). The asymptotic behavior $T_3 \sim N^3$ has been divided out. The lines correspond to power laws $T_{3} \sim N^{3.2}$ (full line) or $T_{3} \sim N^{3.5}$ (broken line). The short dashed line gives the asymptotic limit.}
\end{center}
\end{figure}
\newpage

\begin{figure}
\label{fig6}
\begin{center}
\epsfig{figure=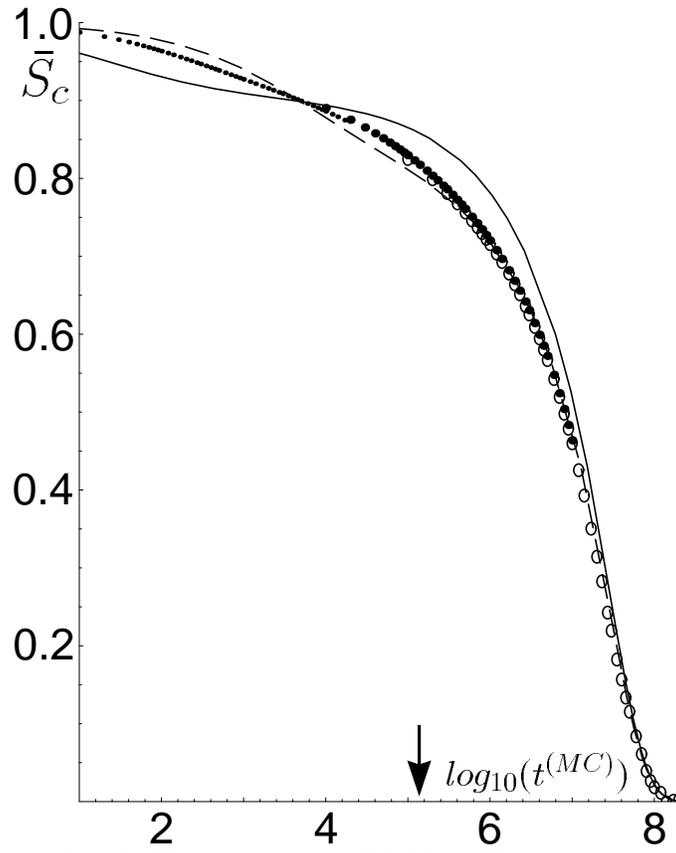,width=0.5\linewidth} 
\caption[]{Fit of $\bar{S}_{dG}$ (Eq.\ (3.9)) to data for 
$N = 320,~ q = 1.0$. Solid line: $N_{e} = 3.69$, corresponding to $B_{dG} = 0.903$; dashes: $B_{dG} = 0.82,~ t_{1}$ (Eq.\ (3.8)) divided 
by $200$. The arrow points to the equilibration time $T_{2}^{(MC)}$. 
Data, small dots: short time runs $(t_{\rm max}^{(MC)} = 10^{5})$; 
heavy dots: medium time runs $(t_{\rm max}^{(MC)} = 10^{8})$, circles: long time runs $(t^{(MC)}_{\rm max} = 5 \cdot 10^{10})$ .}
\end{center}
\end{figure}
\newpage

\begin{figure}
\label{fig7}
\begin{center}
\epsfig{figure=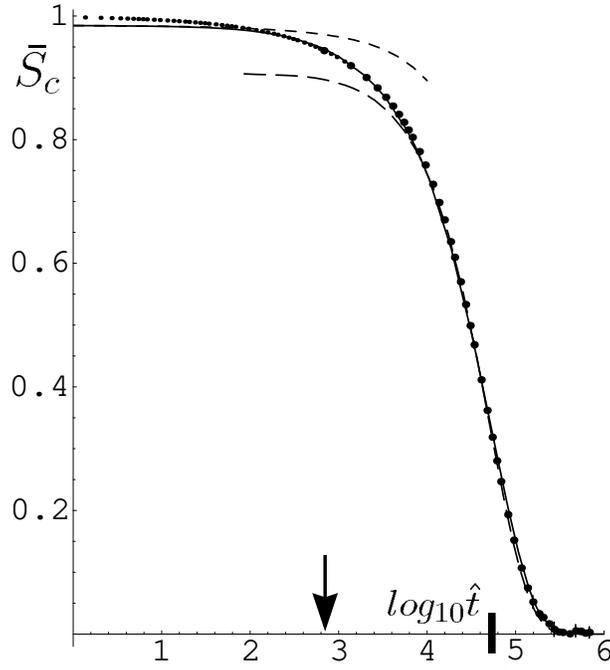,width=0.45\linewidth} 
\caption[]{Monte Carlo data for $\bar{S}_{c}(q,t,N)$ compared to 
reptation theory (full line), as function of $\log_{10} \hat{t}$. 
Chain length $N = 80$; wave vector $q = 0.5024~ (q^{2} R_{g}^{2} = 3.32)$. 
Long dashed line: best fit to the primitive chain model $(\bar{S}_{DE})$. 
Short dashed line: $\bar{S}_{c}$ calculated neglecting tube destruction. 
The arrow points to $\hat{T}_{2}$, the heavy slash gives $\hat{T}_{3}$. 
Data, small dots: average over $10^{4}$ short time runs; heavy dots: 
averages over 50 medium time runs.}
\end{center}
\end{figure}
\newpage

\begin{figure}
\label{fig8}
\begin{center}
\epsfig{figure=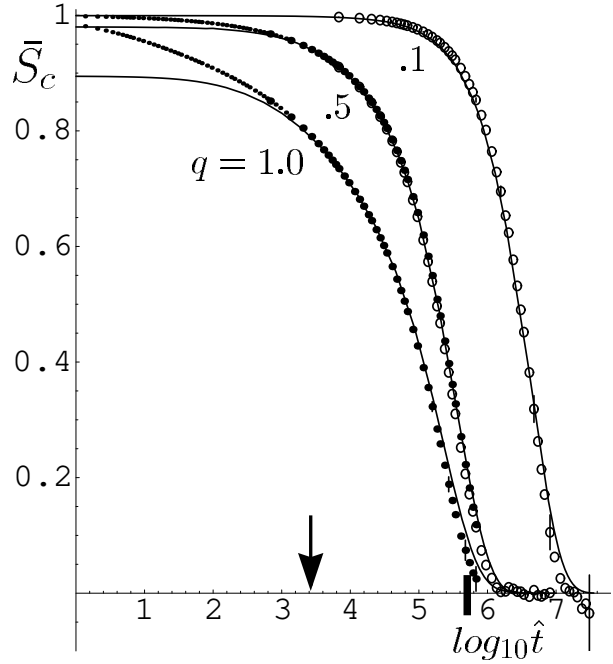,width=0.45\linewidth} 
\caption[]{Data compared to reptation theory. $N = 160$; 
$q = 0.1 ~(q^{2} R_{g}^{2} = 0.267)$; $q = 0.5 ~(q^{2} R_{g}^{2} = 6.665)$; 
$q = 1.0 ~(q^{2} R_{g}^{2} = 26.61)$. Arrow: $\hat{T}_{2}$; 
slash: $\hat{T}_{3}$. Small dots: short time runs; heavy dots: 
runs for intermediate time; circles: long time runs. The time regimes 
overlap. Some error bars (two standard deviations) are also given.}
\end{center}
\end{figure}

\begin{figure}
\label{fig9}
\begin{center}
\epsfig{figure=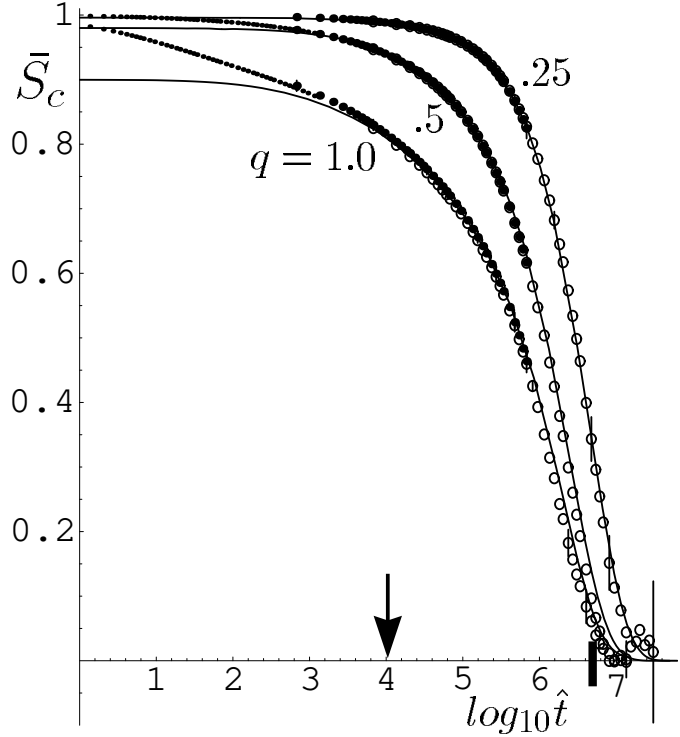,width=0.5\linewidth} 
\caption[]{Same as Fig.~8, but for $N = 320$; 
$q = 0.25 ~(q^{2} R_{g}^{2} = 3.33)$; $q = 0.5~ (q^{2} R_{g}^{2} = 13.33)$; 
$q = 1.0~ (q^{2} R_{g}^{2} = 53.22)$.}
\end{center}
\end{figure}

\begin{figure}
\label{fig0}
\begin{center}
\epsfig{figure=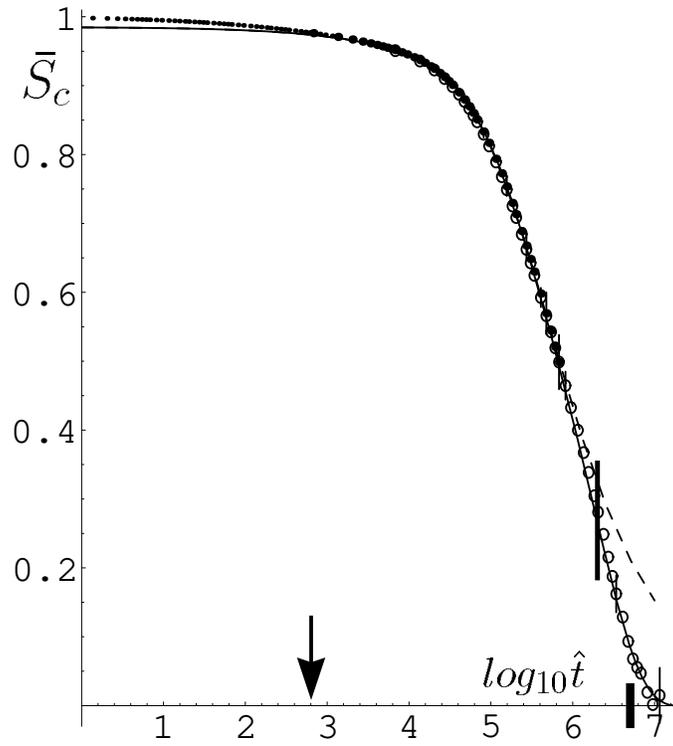,width=0.5\linewidth} 
\caption[]{Coherent structure function of the central piece of length 
$M = 80$ in a chain of length $N = 320$, normalized to the static 
structure function of a chain of length $M$. Wave vector $q = 0.5$. 
Full line: full result of the theory; broken line: result neglecting 
tube destruction. The long slash through the curve at 
$\log_{10} \hat{t} \approx 6.27$ gives the time where tube destruction 
on average reaches the subchain $M = 80$, as derived from our theory. 
Other symbols as in Figs.\ 8 and 9.
}
\end{center}
\end{figure}
\newpage

\begin{figure}
\label{fig11}
\begin{center}
\epsfig{figure=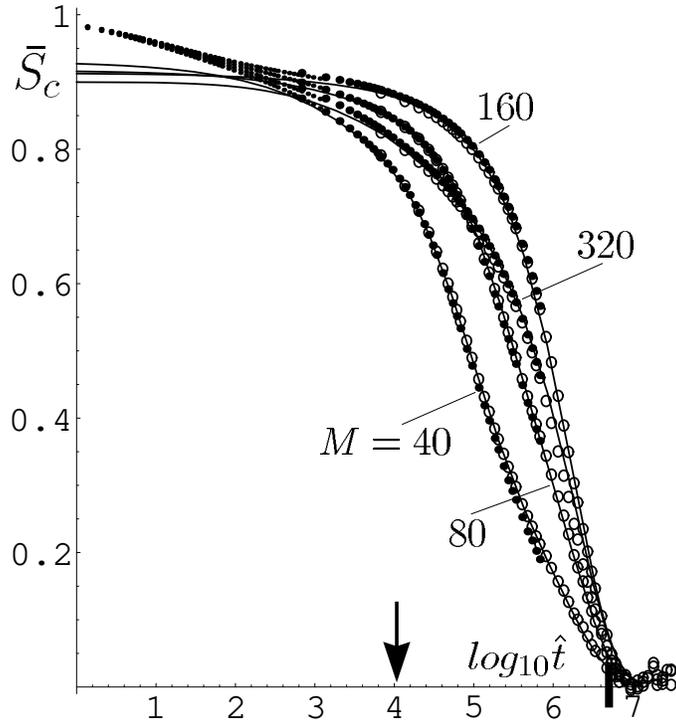,width=0.5\linewidth} 
\caption[]{Normalized coherent structure functions of central subchains 
of lengths $M$ in a chain of length $N = 320$. Wave vector $q = 1.0$. 
$M = 320$ represents scattering from the total chain. 
Symbols are as in Figs.\ 8, 9.}
\end{center}
\end{figure}

\begin{figure}
\label{fig12}
\begin{center}
\epsfig{figure=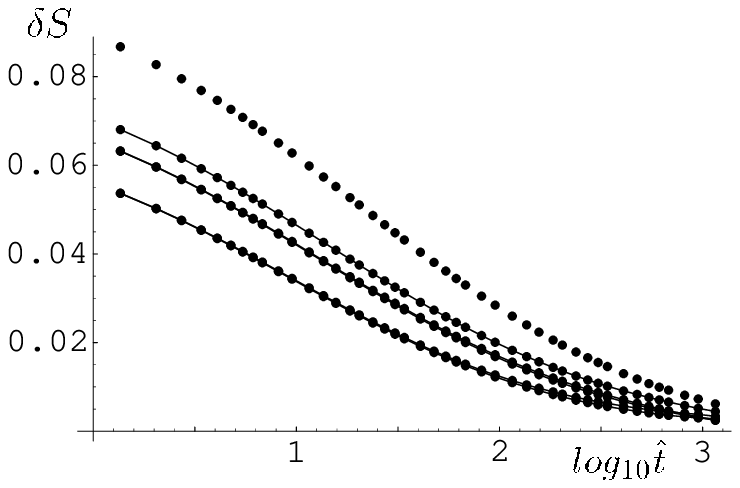,width=0.5\linewidth} 
\caption[]{Deviation between theory and experiment: 
$\delta S = \bar{S}\mbox{(Monte Carlo)} - A \cdot \bar{S}~\mbox{(theory)}$, 
for $q = 1.0$. Connected data points: results for $M = 40~ (N = 160,~320)$, 
$M = 80~ (N = 160,~320)$, $M = 160~(N = 320)$, (from below). 
For $M = 40,~80$ the results for the two different lengths $N$ 
of the total chain fall right on top of each other. 
Disconnected points: $M = N = 160$.}
\end{center}
\end{figure}   


\begin{thebibliography}{88}
\bibitem{Z1} P.G. De Gennes, J. Chem. Phys. {\bf 55}, 572 (1971).
\bibitem{Z2} M. Doi, S.F. Edwards, {\it The Theory of Polymer Dynamics} 
             (Clarendon Press, Oxford, 1986).
\bibitem{Z3} L. Sch\"afer, U. Ebert, A. Baumg\"artner, Phys. Rev. E. {\bf 65}, 
061505 (2002).
\bibitem{Z4} M. Doi, S.F. Edwards, J. Chem. Soc. Faraday Trans. 2,
              {\bf 74}, 1789 (1978).
\bibitem{Z5} P.G. de Gennes, J. Physique {\bf 42}, 735 (1981).                 
      
\bibitem{Z6}  M. Doi, J. Polymer Sci: Polymer Phys. Ed. {\bf 21}, 667 (1983).
\bibitem{Z7} T.P. Lodge, N.A. Rotstein, S. Prager, Advances Chem. Phys. LXXIX,
              ed: Prigogine and Rice  (Wiley 1990).
\bibitem{Z8}  P.G. De Gennes, J. Physique {\bf 36}, 1199 (1975).
\bibitem{Z9} T. Kreer, J. Baschnagel, M. M\"uller, K. Binder, Macromol.
             {\bf 34}, 1105 (2001).
\bibitem{Z10} M. P\"utz, K. Kremer, G.S. Grest, Europhys. Lett.
              {\bf 49}, 735 (2000).           
\bibitem{Z11} R. Graf, A. Heuer, H.W. Spiess, Phys. Rev. Lett. {\bf 80}, 5738 (
1998).
\bibitem{Z12} R. Kimmich, R.-O. Seitter, U. Beginn, M. M\"oller, N. Fatkullin, 
Chem. Phys. Lett. 
              {\bf 307}, 147 (1999).
\bibitem{Z13} W. Hess, Macromol. {\bf 19}, 1395 (1986).              
\bibitem{Z14} K.S. Schweizer, M. Fuchs, G. Szamel, M. Guenza, H. Tang,
             Macromol. Theory Simul. {\bf 6}, 1037 (1997).
\bibitem{Z15} U. Ebert, A. Baumg\"artner, L. Sch\"afer, 
             Phys. Rev. Lett. {\bf 78}, 1592 (1997). 
\bibitem{Z16} U. Ebert, L. Sch\"afer, A. Baumg\"artner, J. Stat. Phys.
             {\bf 90}, 1325 (1998).
\bibitem{Z17} A. Baumg\"artner, U. Ebert, L. Sch\"afer, J. Stat. Phys. 
             {\bf 90}, 1375 (1998).
\bibitem{Z18} K.E. Evans, S.F. Edwards, J. Chem. Soc., Faraday 
              Trans. 2, {\bf 77}, 1891 (1981). 
\bibitem{Z19} L. Sch\"afer, A. Baumg\"artner, U. Ebert,
              Europhys. J.B {\bf 10}, 105 (1999).                              
              
\bibitem{Z20}  R.D. Willmann, J. Chem. Phys. {\bf 116}, 2688 (2002).
\bibitem{Z21} P. Schleger, B. Farago, C. Lartigue, A. Kollmar, D. Richter,
         Phys. Rev. Lett. {\bf 81},
              124 (1998).
\bibitem{Z22} A. Wischnewski, D. Richter, Europhys. Lett. {\bf 52},  719
              (2000).
\bibitem{Z23} A. Wischnewski, M. Monkenbusch, L. Willner, D. Richter,
              A.E. Likhtman, T.C.B. McLeish, B. Farago, Phys. Rev. Lett.
              {\bf 88}, 058301 (2002).        
\bibitem{Z24} M. P\"utz, K. Kremer, G.S. Grest, Europhys. Lett.
              {\bf 52}, 721 (2000).
\bibitem{Z25} F. Spitzer, Principles of Random Walk (Springer, Heidelberg,
              1976).
\end{thebibliography}
\end{document}